\documentclass[twocolumn,3p]{elsarticle}

\usepackage{multirow}
\usepackage[linesnumbered,boxed,ruled,commentsnumbered]{algorithm2e}
\usepackage{balance}
\usepackage{subfig}
\usepackage{url}
\usepackage{amsmath, amsthm}
\usepackage{amssymb}

\usepackage{graphicx} 
\usepackage{epstopdf}

\usepackage[normalem]{ulem}
\useunder{\uline}{\ul}{}

\biboptions{sort&compress}
\journal{Knowledge Based Systems}

\begin{document}

\begin{frontmatter}

\title{A Light Heterogeneous Graph Collaborative Filtering \\ Model using Textual Information}

\author[1]{Chaoyang Wang}
\ead{sunwardtree@hust.edu.cn}

\author[1]{Zhiqiang Guo}
\ead{zhiqiangguo@hust.edu.cn}

\author[2]{Guohui Li}
\ead{guohuili@hust.edu.cn}

\author[1]{Jianjun Li}
\ead{jianjunli@hust.edu.cn}

\author[1]{Peng Pan}
\ead{panpeng@hust.edu.cn}

\author[1]{Ke Liu}
\ead{hustlk@hust.edu.cn}


\address[1]{School of Computer Science and Technology, \\Huazhong University of Science and Technology,  Wuhan, China, 430074}
\address[2]{School of Software Engineering, \\Huazhong University of Science and Technology,  Wuhan, China, 430074}


\begin{abstract}
Due to the development of graph neural networks, graph-based representation learning methods have made great progress in recommender systems. However, data sparsity is still a challenging problem that most graph-based recommendation methods are confronted with. Recent works try to address this problem by utilizing side information. In this paper, we exploit the relevant and easily accessible textual information by advanced natural language processing (NLP) models and propose a light RGCN-based (RGCN, relational graph convolutional network) collaborative filtering method on heterogeneous graphs. Specifically, to incorporate rich textual knowledge, we utilize a pre-trained NLP model to initialize the embeddings of text nodes. Afterward, by performing a simplified RGCN-based node information propagation on the constructed heterogeneous graph, the embeddings of users and items can be adjusted with textual knowledge, which effectively alleviates the negative effects of data sparsity. Moreover, the matching function used by most graph-based representation learning methods is the inner product, which is not appropriate for the obtained embeddings that contain complex semantics. We design a predictive network that combines graph-based representation learning with neural matching function learning, and demonstrate that this architecture can bring a significant performance improvement. Extensive experiments are conducted on three publicly available datasets, and the results verify the superior performance of our method over several baselines.
\end{abstract}


\end{frontmatter}

\section{Introduction}
\label{sec: intro}

In recommender systems, the user-item interactions and the side information about them can be represented as a graph composed by a set of objects~(nodes) and their relationships~(edges).
As these graphs naturally contain at least two different types of nodes, i.e., user $u$ and item $v$, they belong to the heterogeneous graphs (heterographs for short).
Due to the great expressive power of these graphs and the rapid development of graph-based methods~\cite{kipf2016gcn, schlichtkrull2018rgcn, wu2019simplifying, chen2020simple}, graph-based recommendation methods~\cite{wang2019ngcf, chen2020revisiting, he2020lightgcn, wang2018ripplenet, Wang2019kgcn, wang2019kgat, fan2019graph, Liu2020Hetero} have received increasing attention in recent years.

In order to gain more effective representations, some researchers try to exploit the structure of interaction graph by propagating user and item embeddings on it.
For example, inspired by the graph convolutional network~(GCN)~\cite{kipf2016gcn}, Wang \emph{et al.} propose NGCF (neural graph collaborative filtering)~\cite{wang2019ngcf} to derive more effective embeddings.
Later, both LR-GCCF (linear residual graph convolutional collaborative filtering)~\cite{chen2020revisiting} and LightGCN~\cite{he2020lightgcn} find that discarding some components in standard GCN (such as non-linear activation and feature transformation) can yield better performance for collaborative filtering (CF).
Though achieving remarkable progress, existing graph-based representation learning methods still suffer from the data sparsity problem, which is very common in real-world recommendations. Since the abovementioned methods are not designed to address this problem, they may suffer serious performance degradation when facing it. One of the typical solutions to alleviate this problem is adding side information~\cite{GUO2017202}, which can increase indirect node connections or enrich the features of nodes. The recommendation-related textual information contains a wealth of side information, but most of the exiting efforts~\cite{Wang2019kgcn, fan2019graph, wang2019kgat} that introduce the side information ignore it, which limit their performance.

To alleviate the data sparsity problem, similar to HGNR~(heterogeneous graph neural recommendation)~\cite{Liu2020Hetero}, based on the interaction graph (Figure~\ref{fig:graph}-(a)), we introduce the description and comment nodes to construct our heterograph (Figure~\ref{fig:graph}-(b)). But it is worth mentioning that simply adding these nodes cannot increase indirect connections between users and items.
Nevertheless, the text nodes themselves contain plentiful features, which are helpful to adjust the embeddings of users and items.
Inspired by this, we utilize advanced natural language processing~(NLP) techniques to initialize the embeddings of text nodes, hence integrate with the semantics to enhance the representation ability. Through the node information propagation, the embeddings of users and items will contain the textual knowledge, which can  help alleviate the data sparsity problem effectively.

HGNR utilizes standard GCN to learn embeddings on heterograph. In other words, it transforms the heterogeneous learning process to a homogeneous one, which may cause unexpected information loss and performance degradation.
By iteratively applying the GCN framework over different relations,  RGCN (Relational Graph Convolutional Network) has been shown to be capable of dealing with the highly multi-relational data characteristic on heterograph~\cite{schlichtkrull2018rgcn}. In view of this,  applying RGCN to propagate information on heterograph is a better choice. 
However, simply utilizing the standard RGCN is not desirable, since in order to distinguish different relations,  the standard RGCN always has a larger parameter scale than the GCN on the same heterograph, which makes the training of the RGCN-based model difficult and inefficient. To address this problem, we propose a modified RGCN to propagate information on heterograph in this work.
Firstly, it has been proved in \cite{wu2019simplifying, he2020lightgcn} that feature transformation and non-linear activation of the standard GCN contribute little to the CF performance. Similarly, we discard these two components to simplify RGCN and therefore improve its performance and efficiency.
Secondly, we adopt the weighted layer combination in LightGCN~\cite{he2020lightgcn} and the initial residual in GCNII~\cite{chen2020simple} to further modify RGCN for better performance. The experiment results show that the above modifications on RGCN can improve the recommendation performance significantly.

After deriving efficient user and item embeddings via feature propagation of the modified RGCN, another important component of the recommender system is the matching function. Almost all of existing graph-based representation learning methods utilize inner product as the matching function~\cite{wang2019ngcf, he2020lightgcn, Wang2019kgcn, wang2019kgat}.
However, in the heterographs constructed by us, both the description and comment nodes contain rich semantics, and the simple linear inner product is ineffective for these complex features.
To tackle the semantic features learned by GNN from heterographs, GraphRec~\cite{fan2019graph} concatenated latent factors of users and items and utilized an MLP (multi-layer perception) to predict ratings.
Meanwhile, DeepCF~\cite{deng2019deepcf}, which applies MLP neural network on user-item interaction matrix, combines the strengths of neural representation learning and neural matching function learning to overcome the limitations of each other.
Inspired by GraphRec and DeepCF, we design a predictive network that incorporates the graph-based representation learning with the neural matching function learning to better fit the semantics.
Specifically, the combination of graph-based representation learning and the neural matching function learning can capture the low-rank user-item relations and learn the complex matching function jointly.

To sum up, the main contributions of our work in this paper are as follows:
\begin{itemize}
\item We build a heterograph and initialize the embeddings of text nodes with advanced NLP techniques to exploit the knowledge in the related textual information. Then, by utilizing a modified RGCN, the semantics in these text nodes can be propagated to derive the representations of users and items.

\item In view that inner product is ineffective for complex semantics, we propose a predictive framework that combines graph-based representation learning with neural matching function learning for better predictions.

\item Extensive experiments are conducted on three benchmark datasets, the results verify the superior performance of our method over several baselines.
\end{itemize}

The remainder of this paper is organized as follows:  Section~\ref{sec: relatedwork} discusses related work; Section~\ref{sec: method} details the proposed model; Section~\ref{sec: experiment} presents and analyzes the experimental results; Finally, Section~\ref{sec: conclusion} concludes this paper.

\section{Related Work}
\label{sec: relatedwork}

In this section, we review existing related works on graph-based, text-based and CF-based recommendations, respectively.
\subsection{Graph-based Recommendation}
\begin{figure}[t]
	\centering
	\includegraphics[width=0.8\linewidth]{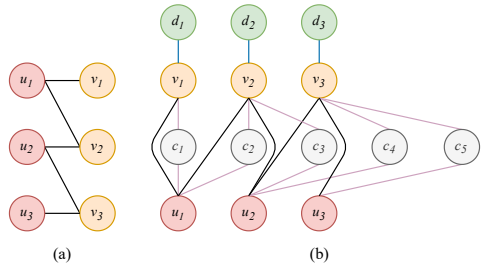}
	\caption{Different graphs. (a) the bipartite graph; (b) the heterograph with description and comment nodes.}
	\label{fig:graph}
\end{figure}

To distinguish with other types of heterographs, the heterographs that only contain user-item interactions are a kind of bipartite graphs, as depicted in Figure~\ref{fig:graph}~(a).
For bipartite graphs, early works, such as ItemRank~\cite{gori2007itemrank} and BiRank~\cite{he2016birank}, adopt the idea of label propagation to capture the CF effect, but the lack of model parameters for optimization limits their performance.
Recently, HOP-Rec~\cite{yang2018hop} incorporates matrix factorization~(MF) and random walks, but it only uses the graph-info to regularize to the MF and does not contribute to the embeddings.
GC-MC~(graph convolutional matrix completion)~\cite{berg2017gc-mc} applies a GCN-based auto-encoder framework on the bipartite user-item graph, but it only employs GCN for link prediction between users and items.
Inspired by GCN, NGCF~\cite{wang2019ngcf} exploits the collaborative signal in the embedding function and explicitly encodes the signal in the form of high-order connectivity by performing embedding propagation.
The embedding propagation rule of NGCF is the same as that of standard GCN, including feature transformation, neighborhood aggregation, and non-linear activation.
But GCN is originally proposed for node classification on the attributed graph, where each node has rich attributes, whereas in the bipartite graph used by NGCF, each node is only described by a one-hot ID. By removing feature transformation and non-linear activation that will negatively increase the difficulty for training in NGCF, LightGCN~\cite{he2020lightgcn} achieves significant accuracy improvements.
However, as mentioned before, these graph-based representation learning methods may suffer serious performance degradation when facing the data sparsity problem on bipartite graphs.

\setlength{\tabcolsep}{10pt}
\begin{table*}[t]
\centering
\small
\caption{A summarization of the properties to the most related works.}
\vspace{-1em}
\label{tab:relate-works}
\linespread{1.1}\selectfont
\begin{tabular}{c|ccl}
\hline
\textbf{Algorithm}        & \textbf{Based Model}      & \textbf{Data}                    & \multicolumn{1}{c}{\textbf{Property}}               \\
\hline
NGCF~\cite{wang2019ngcf}                      & GCN                       & Bipartite Graph                  & Inner product, Standard GCN.          \\
\multirow{2}{*}{LightGCN~\cite{he2020lightgcn}} & \multirow{2}{*}{GCN}      & \multirow{2}{*}{Bipartite Graph} & Inner product,  Standard GCN without non-linear        \\
                          &                           &                                  & activation and feature transformation.  \\
\hline
KGAT~\cite{wang2019kgat}                     & GAT                       & Heterograph                      & Inner product, Knowladge graph.       \\
\multirow{2}{*}{GraphRec~\cite{fan2019graph}} & \multirow{2}{*}{GNN, DNN} & \multirow{2}{*}{Graphs}          & Deep matching function learning,     \\
                          &                           &                                  & Attention mechanism, Social   information.           \\
\hline
\multirow{2}{*}{HGNR~\cite{Liu2020Hetero}}     & \multirow{2}{*}{GCN}      & \multirow{2}{*}{Heterograph}     & Inner product, Textual information,   \\
                          &                           &                                  & Utilizing SBERT~\cite{reimers2019sbert}.                \\
\multirow{2}{*}{ANR~\cite{chin2018anr}}      & \multirow{2}{*}{DNN}      & \multirow{2}{*}{Matrix}          & Aspect importance product, Attention mechanism, \\
                          &                           &                                  & Textual information.
\\
\hline
NGPR~\cite{hu2021neural}                      & GCN, DNN                  & Bipartite Graph                  & Deep matching function learning.      \\
\multirow{2}{*}{NeuMF~\cite{he2017ncf}}    & \multirow{2}{*}{DNN}      & \multirow{2}{*}{Matrix}          & Deep matching function, Linear representation    \\
                          &                           &                                  & learning.
\\
\multirow{2}{*}{DeepCF~\cite{deng2019deepcf}}   & \multirow{2}{*}{DNN}      & \multirow{2}{*}{Matrix}          & Deep matching function, Deep representation      \\
                          &                           &                                  & learning. 
\\
\hline
\end{tabular}
\end{table*}


To alternative the data sparsity problem, existing works~\cite{Wang2019kgcn, fan2019graph, wang2019kgat, Liu2020Hetero} based on graph neural networks (GNNs) generally focus on leveraging side information or auxiliary information, like KG (knowledge graph)~\cite{guo2020survey}.
Some early works only exploit the knowledge of KG to enrich the representations~\cite{zhang2016cke, wang2018dkn} or only leverage the connectivity patterns of the entity in KG for better performance~\cite{sun2018recurrent, wang2019explainable}.
To fully exploit the knowledge of KG, RippleNet~\cite{wang2018ripplenet} stimulates the propagation of user preferences by automatically and iteratively extending a user’s potential interests along with links in KG.
KGCN~(knowledge graph convolutional networks)~\cite{Wang2019kgcn} utilizes GCN to effectively mine their representations that contains semantic information of KG and users' personalized interests in relations.
When calculating the representation of a given entity in KG, GCN-based KGCN aggregates and incorporates neighborhood information with bias.
KGAT~(knowledge graph attention network)~\cite{wang2019kgat} refines a node's embedding by propagating the embeddings from its neighbors in KG and employs an attention mechanism to discriminate the importance of its neighbors.
For the social recommendation, Fan \emph{et al.} propose GraphRec~\cite{fan2019graph}, which coherently models social graph and user-item graph and utilizes an MLP to predict ratings.
However, these aforementioned methods just focus on increasing indirect connections for user and item nodes, but ignore the plentiful knowledge contained in the textual information (such as descriptions and comments).
Moreover, compared with constructing KGs, directly using relevant and easily available textual information is more convenient and low-cost.
Liu \emph{et al.} propose HGNR~\cite{Liu2020Hetero} to learn the embeddings of users and items by using the GCN on a heterograph, which is constructed from user-item interactions, social links, and semantic links predicted from the social network and textual comments.
However, HGNR just utilizes the standard GCN but does not consider the difference between nodes.
Therefore, the learning of heterograph degenerates as a homogeneous process, which may cause unexpected information loss.
Moreover, the HGNR utilizes the simple inner product as the matching function for the embeddings that contain complex semantics, which also limits its performance.

\subsection{Text-based Recommendation}
The application of deep learning methods in NLP~\cite{Mikolov2013w2v, pennington2014glove, devlin2019bert} makes it possible to incorporate textual information that human beings can understand with recommendation methods.
Among the textual information in recommendations, the comments which contain users' attitudes, and descriptions that contain items' attributes, are vitally important.
There are works that utilize sentiment analysis~\cite{qiu2016aspect, Bauman2017}, convolutional neural networks~\cite{deng2018neural} and pre-trained word vectors on large corpora~\cite{Mikolov2013w2v, chin2018anr} to get embeddings of comments and descriptions.
Specially, some researchers are trying to combine KG embedding models with the textual information of entities. For example, Richard \emph{et al.}~\cite{socher2013reasoning} introduce an expressive neural tensor network suitable for reasoning over relationships between two entities, and find that the performance is improved when entities are represented as an average of the word vectors.
Xie \emph{et al.}~\cite{xie2016represent} propose an RL method for KGs by taking advantage of entity descriptions, which are leaned by a continuous bag-of-words model and deep convolutional neural model.
Xiao \emph{et al.}~\cite{xiao2017ssp} propose the semantic space projection model, which jointly learns from the symbolic triples and textual descriptions.
The users' and items' features contain in these embeddings can alleviate the negative effects of data sparsity, and thus improve the recommendation performance.
However, the NLP techniques used by these text-based methods have weak context awareness, which may limit their ability to understand the textual information.

The  state-of-the-art contextual NLP techniques, such as Transformer~\cite{vaswani2017transformer} and bidirectional encoder representations from transformers~(BERT)~\cite{devlin2019bert}, improve the performance of many NLP tasks significantly.
Some researchers have applied these pre-trained NLP models in citation recommendations~\cite{hassan2019bert, jeong2019context}.
Hebatallah \emph{et al.}~\cite{hassan2019bert} show that the integration of traditional methods with pre-trained sentence encoders can retrieve more relevant papers.
Chanwoo \emph{et al.}~\cite{jeong2019context} combine the context embeddings encoded by pre-trained BERT and the document embeddings encoded by GCN to learn the proper references for scientific papers.
For the item recommendations, on the heterograph that contains users, items and comments, HGNR~\cite{Liu2020Hetero} explores the potential links between users and items by utilizing SBERT~\cite{reimers2019sbert} to enrich features for user and item nodes.
In this work, to figure out the influence of textual context, we exploit both non-contextual GloVe~\cite{pennington2014glove} and contextual SBERT to gain the initial embeddings of text nodes.

\subsection{CF-based Recommendation}
Recently, DNNs that can capture high-order user-item relationships and enable the codification of more complex abstractions have changed recommendation architecture dramatically~\cite{zhang2019deep}.
Depending on the focus of the CF-based DNN methods, we divide these methods into three categories~\cite{deng2019deepcf}: representation learning, matching function learning, and the combination of them.

The main idea of representation learning for recommendation is to map users and items in a common representation space where they can be compared directly.
By using a two pathway neural network representation learning architecture, deep matrix factorization~(DMF)~\cite{xue2017dmf} maps the users and items into a common low-dimensional space with non-linear projections, and then utilizes cosine similarity as the matching function to calculate predictive scores.
Similar to DMF, to obtain matching scores, most of the graph-based representation learning methods~\cite{wang2019ngcf, chen2020revisiting, he2020lightgcn, Wang2019kgcn, wang2019kgat} only utilize the inner product as the matching function.
For these representation learning methods, the matching function used in the predictive part is too simple to deal with the complex features and cannot fit the complex score distribution well.

Matching function learning is aiming to learn the complex matching function to model latent features by utilizing DNNs.
He \emph{et al.}~\cite{he2017ncf} figure out that linear inner product limited the ability to learn the complex interaction structure. To learn the complex structure of user interaction data, they replace the inner product matching function with a non-linear MLP architecture.
Moreover, by fusing the neural matching function learning structure MLP with a representation learning structure generalized matrix factorization, NeuMF is proposed to obtain better performance.
Later, Chen \emph{et al.} propose J-NCF~\cite{chen2019joint}, which applies a joint neural network that couples deep feature learning (representation learning) and deep interaction modeling (matching function learning) with a rating matrix.
Further, considering the DNN-based representation learning and matching function learning suffered from two fundamental flaws, i.e., the limited expressiveness of inner product and the weakness in capturing low-rank relations respectively, Deng \emph{et al.}~\cite{deng2019deepcf} propose DeepCF, which combines the strengths of neural representation learning and neural matching function learning, to overcome such flaws.
Most recently, based on a bipartite graph, NGPR~(neural graph personalized ranking)~\cite{hu2021neural} is proposed to capture the user-item geometric structure by exploiting rich complementary embedding propagation and endowing the interaction modeling with neural collaborative signals into embeddings by an MLP.

Inspired by the success of NGPR and DeepCF, to better exploit textual features and learn a complex matching function, we design a new framework that combines graph-based representation learning with neural matching function learning.
Different from the predictive part of NGPR, our predictive network adds a neural representation learning part to capture the low-rank user-item relations.
As for DeepCF, it is a neural network-based method with the input of interaction matrices, which is different from our method.

In order to better illustrate the differences between the most related works with ours, we provide a summarization table to list their properties, as shown in Table~\ref{tab:relate-works}.

\section{Proposed Method}
\label{sec: method}

\subsection{Framework Overview}
We consider a recommender system with $M$ users $\mathrm{U}=\{u_1, \ldots, u_M\}$, $N$ items $\mathrm{V}=\{v_1, \ldots, v_N\}$, $P$ descriptions $\mathrm{D}=\{d_1, \ldots, d_P\}$ and $Q$ comments $\mathrm{C}=\{c_1, \ldots, c_Q\}$, and the hidden state of them in $l$-th propagation layer are denoted as $\textbf{\textrm{U}}^{(l)}=\{\textbf{\textrm{u}}^{(l)}_1, \ldots, \textbf{\textrm{u}}^{(l)}_M\}$, $\textbf{\textrm{V}}^{(l)}=\{\textbf{\textrm{v}}^{(l)}_1, \ldots, \textbf{\textrm{v}}^{(l)}_N\}$, $\textbf{\textrm{D}}^{(l)}=\{\textbf{\textrm{d}}^{(l)}_1, \ldots, \textbf{\textrm{d}}^{(l)}_P\}$ and $\textbf{\textrm{C}}^{(l)}=\{\textbf{\textrm{c}}^{(l)}_1, \ldots, \textbf{\textrm{c}}^{(l)}_Q\}$, respectively.
$Y \in \mathbb{R}^{M \times N}$ denotes the implicit feedback matrix, where $y_{m,n}$ is the implicit feedback of user $u_m$ on item $v_n$. The value of $y_{m,n}$ can be defined as:
\begin{equation}
y_{m,n}=\left\{
\begin{array}{ll}
1, & \text{if}~~ u_m \text{ has interacted with } v_n; \\
0, & \text{otherwise.}
\end{array}
\right.
\label{eq:1}
\end{equation}

\begin{figure}[t]
	\centering
	\includegraphics[width=1\linewidth]{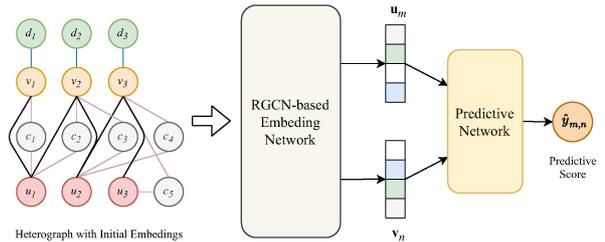}
	\caption{Overview of LT-HGCF~(An example of $u_m$, $v_n$).}
	\label{fig:frame}
\end{figure}

Figure~\ref{fig:frame} gives an overview of the proposed Light Text-based Heterogeneous Graph Collaborative Filtering~(LT-HGCF). Firstly, in the heterograph, we initialize $\textbf{\textrm{u}}^{(0)}_m$ and $\textbf{\textrm{v}}^{(0)}_n$ with zero vectors, and initialize $\textbf{\textrm{d}}^{(0)}_p$ and $\textbf{\textrm{c}}^{(0)}_q$ with existing NLP techniques. Specifically, to figure out the impact of textual context, we utilize pre-trained non-contextual GloVe~\cite{pennington2014glove} and contextual SBERT~\cite{reimers2019sbert}, respectively.
Then, by propagating information on the RGCN-based embedding network, the textual knowledge in the text node embeddings can be integrated into the embeddings of users and items to obtain the final weighted layer combination representations $\textbf{\textrm{u}}_m$ and $\textbf{\textrm{v}}_n$ for user $u_m$ and item $v_n$.
Finally, the predictive network, which combines graph-based representation learning with neural matching function learning, takes $\textbf{\textrm{u}}_m$ and $\textbf{\textrm{v}}_n$ as input to get the predictive scores.
In the training phase, we utilize pairwise BPR~(bayesian personalized ranking) loss~\cite{rendle2009bpr} to learn the parameters of the embedding network and the predictive network jointly.

\subsection{Embeddings Initialization of Text Nodes}
Textual information contains rich knowledge that can alleviate the negative effects of data sparsity~\cite{zheng2017joint}.
As mentioned above, we utilize two types of NLP techniques, non-contextual GloVe~\cite{pennington2014glove} and contextual SBERT~\cite{reimers2019sbert}, to initialize the embeddings of text nodes respectively.

\begin{figure*}[t]
	\centering
	\includegraphics[width=0.85\linewidth]{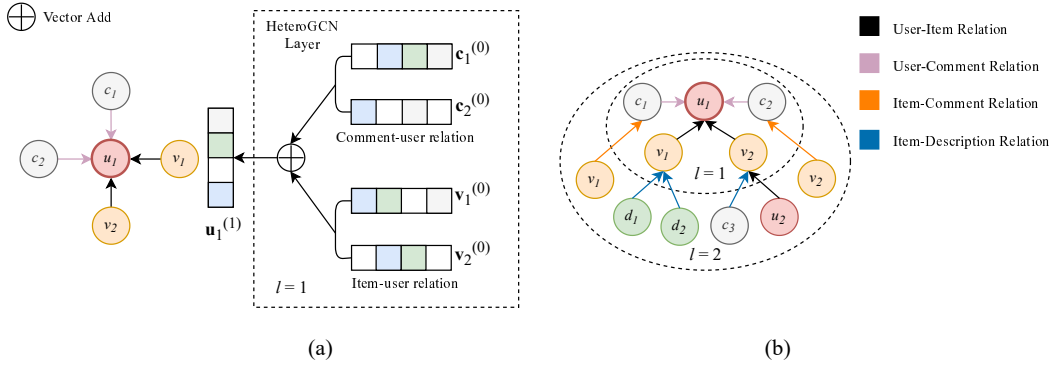}
	\caption{RGCN-based embedding network \emph{w.r.t.} (a) the 1st order HeteroGCN layer; (b) the 1st and 2nd order connectivity (the different line colors represent different relation types and the array direction represents the direction of information aggregation).}
	\label{fig:gcn-net}
\end{figure*}

As for the  non-contextual word-to-vector method GloVe, descriptions and comments contain many meaningless words that can affect the quality of embedding construction, we remove them in advance by comparing with Long Stopword List\footnote{https://www.ranks.nl/stopwords}. Then, we pick up GloVe.6B\footnote{http://nlp.stanford.edu/data/glove.6B.zip}, the pre-trained word vectors on large corpora~(Wikipedia 2014 and Gigaword 5), to calculate the initial embeddings $\textbf{\textrm{d}}^{(0)}_p$ and $\textbf{\textrm{c}}^{(0)}_q$. Specifically,
\begin{equation}
\textbf{\textrm{d}}^{(0)}_p = \frac{1}{n_d}\sum\nolimits_{i=1}^{n_d}\textbf{\textrm{w}}_i, \quad
\textbf{\textrm{c}}^{(0)}_q = \frac{1}{n_c}\sum\nolimits_{i=1}^{n_c}\textbf{\textrm{w}}_i
\label{eq:2}
\end{equation}
where $\textbf{\textrm{w}}_i$ denotes the vector of word ${w}_{i}$, $n_d$~($n_c$) denotes the number of words that $d_p$~($c_q$) contains after removing the stop words.

To get the independent contextual sentence embeddings, there are researches that feed sentences into BERT and then derive a fixed-sized vector by averaging the outputs. However, this common practice yields rather bad sentence embeddings, often worse than averaging GloVe embeddings~\cite{reimers2019sbert}.
To address this problem, Reimers \emph{et al.}~\cite{reimers2019sbert} proposed a contextual sentence embedding method SBERT, which adds a pooling operation after the pre-trained BERT~\cite{devlin2019bert} and utilizes siamese and triplet networks~\cite{schroff2015facenet} to fine-tune the weights of BERT, and has been proved to be capable of producing semantically meaningful sentence embeddings.

In our implementation, considering that the maximum dimension of pre-trained word vectors in GloVe.6B is $300$, which is much smaller than the dimension of BERT-Base~(768), we utilize the pre-trained BERT-Mini~($256$)~\cite{turc2019small_bert} as the target fine-tune model. Moreover, since the MEAN strategy pooling operation~(compute the mean of all output vectors) can always gain the best performance on several NLP tasks, the mean strategy is applied in the SBERT we utilized.
Based on the pre-trained BERT-Mini and mean strategy pooling operation, SBERT can encode the sentences in $d_p$ and $c_q$ to initialize $\textbf{\textrm{d}}^{(0)}_p$ and $\textbf{\textrm{c}}^{(0)}_q$. Note that,  Equation~(\ref{eq:2}) is only for the word-to-vector GloVe, while contextual SBERT can directly encode each sentence of per description or comment.
To avoid the extra effect, we utilize the zero initialization of users embedings $\textbf{\textrm{U}}^{(0)}$ and items embeddings $\textbf{\textrm{V}}^{(0)}$, rather than the random initialization that has been employed in NGCF~\cite{wang2019ngcf} and LightGCN~\cite{he2020lightgcn}.

\subsection{RGCN-based Embedding Network}
To utilize the knowledge in textual information, different from the standard GCN that randomly initializes the embeddings of all the nodes, we initialize the embeddings of the text nodes in our heterograph by advanced NLP techniques.
After that, as shown in Figure~\ref{fig:gcn-net}, we take user $u_1$ as an example to concretely describe the first-order RGCN-based embedding propagation, and then expand it to any nodes in high-order. Note that, since the edges in our heterograph only represent related associations, we do not learn and use the embeddings of them.

\begin{figure*}[t]
	\centering
	\includegraphics[width=0.98\linewidth]{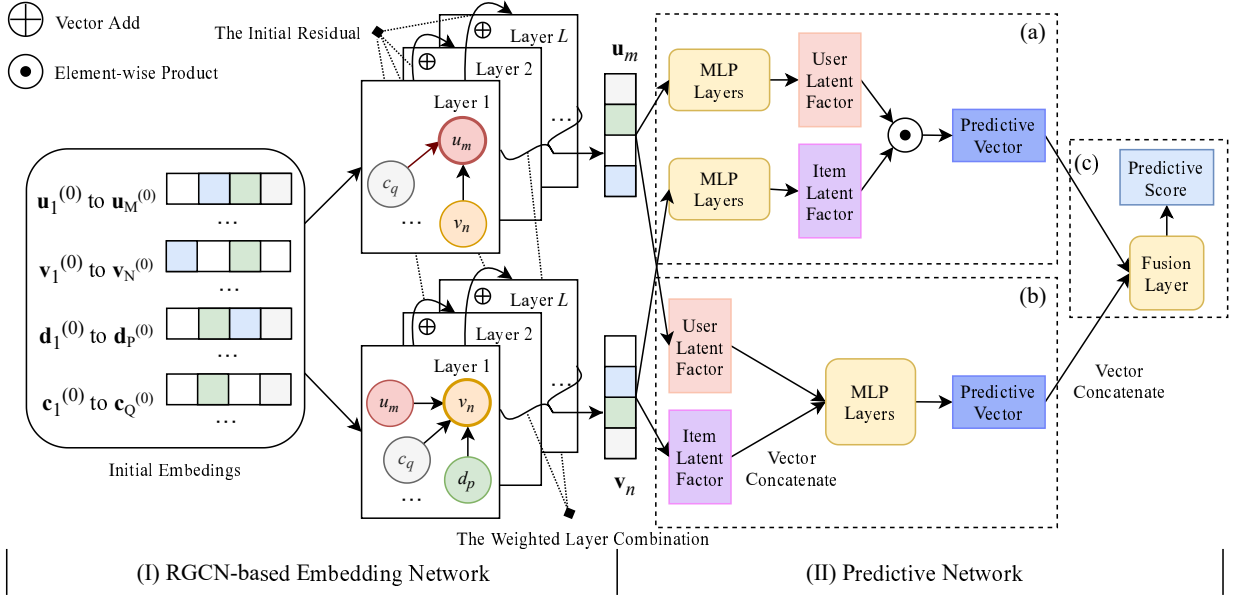}
	\caption{The Network Framework of LT-HGCF \emph{w.r.t.} (a) representation learning part; (b) matching function learning part; (c) fusion part.}
	\label{fig:joint-net}
\end{figure*}

We utilize $e \in \mathrm{E}$ to denote the entities in the heterograph, where $\mathrm{E}=\mathrm{U}\cup\mathrm{V}\cup\mathrm{D}\cup\mathrm{C}$, and use \textbf{\textrm{e}} to denote the representation of $e$. According to feature propagation rule of RGCN~\cite{schlichtkrull2018rgcn}, we formulate message-propagation for user $u_1$ under the relation set $\mathcal{R}$ as:
\begin{equation}
\footnotesize
\textbf{\textrm{u}}^{(l)}_1=\left\{
\begin{array}{ll}
\sum\limits_{r \in \mathrm{\mathcal{R}}} \sum\limits_{m \in \mathrm{\mathcal{N}^r_{u_1}}} g_{m}( \textbf{\textrm{u}}^{(0)}_1, \textbf{\textrm{e}}^{(0)}_j ), & \text{if}~~ l = 1; \\
\sum\limits_{r \in \mathrm{\mathcal{R}}} \sum\limits_{m \in \mathrm{\mathcal{N}^r_{u_1}}} g_{m}( \textbf{\textrm{u}}^{(l-1)}_1, \textbf{\textrm{e}}^{(l-1)}_j ) + \textbf{\textrm{u}}^{(1)}_1, & \text{else.}
\end{array}
\right.
\label{eq:3}
\end{equation}
where $g_{m}(\cdot)$ is the message construction function, and $\mathcal{N}^r_{u_1}$ denotes the set of incoming messages for node $u_1$ under relation $r \in \mathrm{\mathcal{R}}$.
As mentioned above, similar to SGC~\cite{wu2019simplifying} and LightGCN~\cite{he2020lightgcn}, we abandon the non-linear activation function to avoid the unnecessary complexity. Moreover, we apply a variation of the initial residual here.
In GCNII~\cite{chen2020simple}, at each layer, initial residual constructs a skip connection from the input layer. However, the initial embeddings of users and items are zero in our model, which is less helpful for emphasizing the initial knowledge and alleviating the over-smoothing problem.
Therefore, we construct a skip connection from the first layer to other higher layers by directly adding $\textbf{\textrm{u}}^{(1)}_1$, as shown in the skip arrows between layers in the embedding network of Figure~\ref{fig:joint-net}, and is formulated as Equation~(\ref{eq:3}).

In the standard RGCN, $g_{m}(\cdot)$ is typically chosen to be a simple linear transformation with a normalization, i.e., $g_{m}\left( \textbf{\textrm{u}}_1, \textbf{\textrm{e}}_j \right)=\mathcal{L}_{u_1, e_j} W \textbf{\textrm{e}}_j$, where $W$ denotes the weights for feature transformation and $\mathcal{L}_{u_1, e_j} = 1 / \sqrt{|\mathcal{N}_{u_1}||\mathcal{N}_{e_j}|}$ is the symmetric normalization term.
Through solid experiments, \cite{he2020lightgcn} has demonstrated that feature transformation, i.e., the weight matrices in standard GCN, contributes little to the CF performance and even degrades performance.
Therefore, we also discard the weight matrices, and the message construction function in our model is formulated as $g_{m}\left( \textbf{\textrm{u}}_1, \textbf{\textrm{e}}_j \right)=\mathcal{L}_{u_1, e_j} \textbf{\textrm{e}}_j$.

As shown in Figure~\ref{fig:gcn-net}-(a), the first propagation embedding of $u_1$ can be obtained by propagating the hidden state through first-order connectivity, $\textbf{\textrm{u}}^{(1)}_1$.
To gain the high-order connectivity information, we can stack more embedding propagation layers.
As Figure~\ref{fig:gcn-net}-(b) shows, when the number of the propagation layers $l$ increases from $l=1$ to $l=2$, the nodes that can be used to calculate the embedding of $u_1$ increases four~($d_1, d_2, c_3, u_2$). Note that, for an interacted user and item pair in our heterograph, there are two ways to connect them, one is the direct connection~(black arrow in Figure~\ref{fig:gcn-net}-(b)), the other is the indirect connection through the transmission of a comment~(purple arrow in Figure~\ref{fig:gcn-net}-(b)). Therefore, the heterograph we used is a cyclic graph.
But since the RGCN-based method can learn on the cyclic graph, it has no impact on our implementation.

We extend the embedding calculation of $\textbf{\textrm{u}}^{(l)}_1$ to $\textbf{\textrm{e}}^{(l)}_i$, and use $\mathcal{N}^r_{e_i}$ to denote  the set of neighbor entities that directly connected to $e_i$ under relation $r$.
Since the layer combination operation that will be introduced next can capture the same effect as self-connection~\cite{he2020lightgcn}, the same as LightGCN, we remove the self-connection operation in our model. In this way,  an entity $e_i$ is capable to receive the messages propagated from its $l$-hop neighbors by stacking $l$ embedding propagation layers. In the $l$-th step, the embedding of $e_i$ is recursively formulated as,
\begin{equation}
\textbf{\textrm{e}}^{(l)}_i=\left\{
\begin{array}{ll}
\sum\limits_{r \in \mathrm{\mathcal{R}}} \sum\limits_{e_j \in \mathrm{\mathcal{N}^r_{e_i}}} \mathcal{L}_{e_i, e_j} \textbf{\textrm{e}}^{(0)}_j, & \text{if}~~ l = 1; \\
\sum\limits_{r \in \mathrm{\mathcal{R}}} \sum\limits_{e_j \in \mathrm{\mathcal{N}^r_{e_i}}} \mathcal{L}_{e_i, e_j} \textbf{\textrm{e}}^{(l-1)}_j + \textbf{\textrm{e}}^{(1)}_i, & \text{else.}
\end{array}
\right.
\label{eq:4}
\end{equation}
where $\mathcal{L}_{e_i, e_j} = 1 / \sqrt{|\mathcal{N}_{e_i}||\mathcal{N}_{e_j}|}$ is the Laplacian matrix (note that, each type of relation $r$ corresponds to a Laplacian matrix), and $\textbf{\textrm{e}}^{(l-1)}_i$ and $\textbf{\textrm{e}}^{(l-1)}_j$ respectively denote the representations of $e_i$ and $e_j$ generated from the previous message-propagation steps. We have  implemented our model with the matrix form propagation rule as in~\cite{wang2019ngcf,he2020lightgcn}, by which we can simultaneously update the representations of all nodes in a rather efficient way.

After propagating on $l$ layers for all relations, we can obtain the embeddings for all the nodes,  denoted as $\textbf{\textrm{E}}^{(l)}$.
Then, we combine all the layers' embeddings to get the final representations of all nodes $\textbf{\textrm{E}}$,  formally,
\begin{equation}
\textbf{\textrm{E}}=\alpha_{0} \textbf{\textrm{E}}^{(0)} + \alpha_{1} \textbf{\textrm{E}}^{(1)}+\cdots+\alpha_{L} \textbf{\textrm{E}}^{(L)}
\label{eq:5}
\end{equation}
where $\alpha_i$ denotes the weight for the $i$-th layer.
Lastly, taking user $u_m$ and item $v_n$ as an example, the propagation process over $L$ layers among different nodes is shown as the RGCN-based embedding network in Figure~\ref{fig:joint-net}. Based on the initial embeddings of text nodes, the RGCN-based embedding network finally encodes the users and items with rich semantics.

\subsection{Predictive Network}
After propagating with $L$ layers, we obtain the final embeddings $\textbf{\textrm{u}}_m$ and $\textbf{\textrm{v}}_n$ as the input of the predictive part.
Different from the inner product used by most of graph-based representation learning methods, we design a predictive network, which incorporates graph-based representation learning with neural matching function learning.
As shown in Figure 4, the representation learning part is (a), the matching function learning part is (b) and the fusion part is (c).
The relationship of these three parts can be described as follows:  Part (a) and part (b) simultaneously take the outputs of the embedding network as input.
Then, part (a) is used to capture the low-rank user-item relations, and part (b) is used to learn the complex matching function.
Afterward, the output vectors of the above two parts are concatenated and is fed into part (c) to get the final predictive score.

The RGCN-based embedding network, itself belongs to representation learning, the representation learning part in our predictive network is mainly used to adjust the obtained embeddings.
We adopt MLP layers to transform the latent representations for users and items. Specifically, for user $u_m$,
\begin{equation}
\begin{aligned}
\mathbf{h}_{u_m}^{(1)}&=\mathbf{W}_{rl}^{(1)} \textbf{\textrm{u}}_m + \mathbf{b}_{rl}^{(1)} \\
& \cdots \cdots \\
\mathbf{h}_{u_m}^{(X_{rl})}&=\mathbf{W}_{rl}^{(X_{rl})} \mathbf{h}_{u_m}^{(X_{rl}-1)} + \mathbf{b}_{rl}^{(X_{rl})}
\end{aligned}
\label{eq:6}
\end{equation}
where $\mathbf{W}_{rl}^{(x_{rl})}$ and $\mathbf{b}_{rl}^{(x_{rl})}$ denote the weight matrixes and bias vector for the $x_{rl}$-th layer’s perception.
Through sufficient experiments, we find that the activation functions in predictive networks always results in a worse performance. The potential reason is that the activation functions increase the difficulty for learning proper parameters. Thus, we simplify our model by discarding all the activation functions for the abovementioned three parts.
The latent representations $\mathbf{h}_{v_n}^{(X_{rl})}$ for item $v_n$ can be calculated similarity. Then,  the latent predictive vector $\mathbf{h}_{rl}^{(X_{rl})}$ of representation learning is calculated by:
\begin{equation}
\mathbf{h}_{rl}^{(X_{rl})}=\mathbf{h}_{u_m}^{(X_{rl})} \odot \mathbf{h}_{v_n}^{(X_{rl})}
\label{eq:7}
\end{equation}
where $\odot$ denotes the element-wise product.

To better match the complex score distribution, matching function learning replaces the inner product with a neural network.
By taking $\textbf{\textrm{u}}_m$ and $\textbf{\textrm{v}}_n$ as the input, we also adopt MLP layers to learn the matching function.
Therefore, the calculation of latent predictive vector $\mathbf{h}_{ml}^{(X_{ml})}$ is formulated as:
\begin{equation}
\begin{aligned}
\mathbf{h}_{ml}^{(0)}&=c(\textbf{\textrm{u}}_m, \textbf{\textrm{v}}_n) = \left[\begin{array}{l}
\textbf{\textrm{u}}_m \\ \textbf{\textrm{v}}_n \end{array}\right]\\
\mathbf{h}_{ml}^{(1)}&=\mathbf{W}_{ml}^{(1)} \mathbf{h}_{ml}^{(0)}+\mathbf{b}_{ml}^{(1)} \\
& \cdots \cdots \\
\mathbf{h}_{ml}^{(X_{ml})}&=\mathbf{W}_{ml}^{(X_{ml})} \mathbf{h}_{ml}^{(X_{ml}-1)}+\mathbf{b}_{ml}^{(X_{ml})}
\end{aligned}
\label{eq:8}
\end{equation}
where  $c(\cdot)$ denotes the vector concatenation, $\mathbf{W}_{ml}^{(x_{ml})}$ and $\mathbf{b}_{ml}^{(x_{ml})}$ denote the weight matrix and bias vector for the $x_{ml}$-th layer, respectively.

The fusion part is utilized to aggregate the latent predictive vector of the above two parts and predict the score. In this part, the latent predictive vector $\mathbf{h}_{rl}^{(X_{rl})}$ and $\mathbf{h}_{ml}^{(X_{ml})}$ are aggregated by a vector concatenation. Finally, an MLP layer is applied as the mapping function to calculate the predictive scores $\hat{Y}$. Specifically,  for user $u_m$ and item $v_n$, the predictive score $\hat{y}_{m,n}$ is calculated by:
\begin{equation}
\hat{y}_{m,n}=\mathbf{W}_{\text {out}} c(\mathbf{h}_{rl}^{(X_{rl})}, \mathbf{h}_{ml}^{(X_{ml})}) + \mathbf{b}_{\text {out}}
\label{eq:9}
\end{equation}
where $\mathbf{W}_{out}$ and $\mathbf{b}_{out}$ respectively denote the weight matrix and bias vector of the fusion layer.

\begin{algorithm}[t]
	\small
	\caption{Learning Algorithm\label{algo: learning}}
	\KwIn{Entitie set of the heterograph: $\mathrm{E}$, textual information set: $\mathrm{T}$, number of embedding propagation layer: $L$, pairwise training data: $\mathrm{O}$.}
	\KwOut{Parameters of graph embedding network and predictive network: $\theta^G, \theta^P$.}
	Initialize $\textbf{\textrm{D}}^{(0)}$ and $\textbf{\textrm{C}}^{(0)}$ with T by SBERT\;
	Initialize $\textbf{\textrm{U}}^{(0)}$ and $\textbf{\textrm{V}}^{(0)}$ with $\textbf{\textrm{0}}$ vectors\;
	\While {not converged}{
		\For {$l = 1$ to $L$}{
			\For {$e_i$ in $\mathrm{E}$}{
				Update $\textbf{\textrm{e}}_i^{(l)}$ by Equation~(\ref{eq:4}) and store it\;	
			}
			//The layer combination operation\;
			Update $\textbf{\textrm{E}}$ by Equation~(\ref{eq:5}) and store it\;
		}
		\For {$(u_m, v_i, v_j)$ in $\mathrm{O}$}{
			Look up $\textbf{\textrm{u}}_m$, $\textbf{\textrm{v}}_i$, $\textbf{\textrm{v}}_j$ from the final embeddings of nodes\;
			Calculate $\hat{y}_{m,i}$, $\hat{y}_{m,j}$ by Equation~(\ref{eq:9}) and store them\;
		}
		Calculate BPR loss by Equation~(\ref{eq:10})\;
		Utilize Adam to optimize $\theta^G$ and $\theta^P$ with BPR loss\;
	}
	\textbf{return} $\theta^G$, $\theta^P$
\end{algorithm}

\subsection{Optimization}
The BPR loss, which has been intensively used in graph-based recommender systems~\cite{wang2019ngcf, chen2020revisiting, he2020lightgcn}, is utilized in our method to optimize the parameters of graph embedding network~($\theta^G$) and predictive network~($\theta^P$). The objective function is defined as:
\begin{equation}
\operatorname{Loss}=\sum_{(m, i, j) \in \mathrm{O}}-\ln a\left(\hat{y}_{m, i}-\hat{y}_{m, j}\right)+\lambda\|\Theta\|_{2}^{2}
\label{eq:10}
\end{equation}
where $\mathrm{O}=\left\{(m, i, j) |(u_m, v_i) \in \mathrm{R}^{+},(u_m, v_j) \in \mathrm{R}^{-}\right\}$ denotes the pairwise training data, $\mathrm{R}^{+}$ indicates the set of interactions that $y_{m,i}=1$, and $\mathrm{R}^{-}$ indicates the set of interactions that $y_{m,j}=0$; $a(\cdot)$ is the activation function and we use Sigmoid here; $\Theta$ denotes all the trainable model parameters, and $\lambda$ controls the $L_2$ regularization strength to prevent over-fitting.
We adopt mini-batch Adam~\cite{kingma2015adam} to optimize the parameters of embedding network and predictive network jointly. The learning algorithm of LT-HGCF is presented in Algorithm~1.

%
\section{Experiments}
\label{sec: experiment}


\setlength{\tabcolsep}{7pt}
\begin{table*}[t]
	\small
	\centering
	\caption{Statistics of datasets.}
	\label{tab:datasets}
	\begin{tabular}{lrrrrrrr}
		\hline
		\textbf{DataSet} & \#\textbf{Users} & \#\textbf{Items} & \#\textbf{Interactions} & \textbf{Sparsity} & \textbf{Size of Des.} & \textbf{Size of Com.}\\
		\hline
		Music & 5,541 & 3,568 & 64,706 & 0.9967 & 2,338 KB & 65,758 KB\\
		Beauty & 22,363 & 12,101 & 198,502 & 0.9993 & 5,735 KB & 83,251 KB\\
		Clothing & 39,387 & 23,033 & 278,677 & 0.9997 & 3,960 KB & 80,208 KB\\
		\hline
	\end{tabular}
\end{table*}

To demonstrate the effectiveness of the proposed method, we first introduce the experimental settings, and then present the experimental results to answer the following research questions:

\begin{itemize}
\item \textbf{RQ1}: How does our method perform as compared with other state-of-the-art CF methods?
\item \textbf{RQ2}: How does the textual information affect the performance of LT-HGCF?
\item \textbf{RQ3}: How does the performance benefit from the modifications on RGCN and the carefully designed structure of the predictive network?
\item \textbf{RQ4}: How do the key hyper-parameters~(e.g., input and output size of embedding network, the dropout rate of the network and the nodes, depth of layers, etc.) affect the performance?
\end{itemize}

Note that, when analyzing one factor, we keep the others fixed. The default hyper-parameters and implementation details for all the datasets are: the embeddings of textual nodes are initialized with the pre-trained SBERT, the input size of embedding network is $256$~(BERT-Mini)~\cite{turc2019small_bert}, the output size of embedding network is $64$, the hidden dimension of embedding network is $128$, the number of embedding propagation layer is $4$, the depth of representation learning network is $1$, the depth of matching function learning network is $2$, the hidden dimension of the representation learning network and matching function learning network are $128$, the dropout rate of the network (including embedding network and predictive network) and the nodes are $0$, the regularization strength $\lambda$ is $1 \times 10^{-4}$, and the learning rate for the joint learning phase is $1 \times 10^{-3}$. Moreover, we utilize the initial residual and the weighted layer combination operation in the default settings, and each weight $\alpha_l$ for layer combination is simply set as $1 / L$. Our model LT-HGCF is implemented based on Pytorch\footnote{https://github.com/pytorch}, and all the codes have been released in the Github\footnote{https://github.com/SunwardTree/LT-HGCF}.

\subsection{Experimental Settings}
\subsubsection{Datasets}
Jure Leskovec \emph{et al.}~\cite{snapnets} collect and categorize a variety of {\it Amazon} products and built several datasets\footnote{http://snap.stanford.edu/data/amazon/productGraph/\\categoryFiles} including ratings, descriptions, and comments. We evaluate our LT-HGCF and other compared methods on three publicly available {\it Amazon}  datasets: Digital Music (\textbf{Music} for short), \textbf{Beauty} and Clothing Shoes and Jewelry (\textbf{Clothing} for short), which all have at least $5$ comments for each product. Among these datasets, from \textbf{Music} to \textbf{Clothing}, the size and the sparsity are increasing. Table~\ref{tab:datasets} shows the statistical details of the datasets.

\setlength{\tabcolsep}{6pt}
\begin{table*}[t]
\footnotesize
\centering
\caption{Performance comparison. The best performance is in boldface and the second is underlined. }
\label{tab:comparison}
\linespread{1.1}\selectfont
\begin{tabular}{c|c|ccccccc|c}
\hline
\multirow{2}[0]{*}{\textbf{Dataset}} & \multirow{2}[0]{*}{\textbf{Metric}} & \multicolumn{7}{c|}{\textbf{Compared Methods}} & \multirow{2}[0]{*}{\textbf{LT-HGCF}} \\
\cline{3-9}&       & \textbf{SVD} & \textbf{MLP} & \textbf{NeuMF} & \textbf{DeepCF} & \textbf{ANR} & \textbf{NGCF$^*$} & \textbf{LightGCN$^*$} &       \\
\hline
\multirow{4}{*}{Music}            & HR@10                            & 0.1045       & 0.5844       & 0.6253         &     0.5896      & 0.3785       & 0.6114        & {\ul 0.6548}      & \textbf{0.7011}                   \\
                                  & NDCG@10                          & 0.0470       & 0.3590       & 0.3969         &     0.3611      & 0.1939       & 0.3845        & {\ul 0.4179}      & \textbf{0.4424}                   \\
                                  & HR@20                            & 0.2099       & 0.7255       & 0.7674         &     0.7419      & 0.5588       & 0.7533        & {\ul 0.7946}      & \textbf{0.8439}                   \\
                                  & NDCG@20                          & 0.0732       & 0.3945       & 0.4315         &     0.3992      & 0.2394       & 0.4182        & {\ul 0.4527}      & \textbf{0.4762}                   \\
\hline
\multirow{4}{*}{Beauty}           & HR@10                            & 0.1011       & 0.3395       & 0.3610         &     0.3513      & 0.3724       & 0.3337        & {\ul 0.4110}      & \textbf{0.4747}                   \\
                                  & NDCG@10                          & 0.0463       & 0.2024       & 0.2240         &     0.2095      & 0.1986       & 0.1983        & {\ul 0.2519}      & \textbf{0.2840}                   \\
                                  & HR@20                            & 0.2004       & 0.4552       & 0.4793         &     0.4659      & {\ul 0.5382} & 0.4489        & 0.5195            & \textbf{0.6068}                   \\
                                  & NDCG@20                          & 0.0712       & 0.2310       & 0.2537         &     0.2386      & 0.2405       & 0.2276        & {\ul 0.2792}      & \textbf{0.3173}                   \\
\hline
\multirow{4}{*}{Clothing}         & HR@10                            & 0.0992       & 0.1593       & 0.1842         &     0.1833      & {\ul 0.3483} & 0.1459        & 0.2073            & \textbf{0.3896}                   \\
                                  & NDCG@10                          & 0.0451       & 0.0844       & 0.0981         &     0.0979      & {\ul 0.1821} & 0.0740        & 0.1182            & \textbf{0.2191}                   \\
                                  & HR@20                            & 0.1993       & 0.2623       & 0.3153         &     0.2939      & {\ul 0.5104} & 0.2499        & 0.3193            & \textbf{0.5426}                   \\
                                  & NDCG@20                          & 0.0701       & 0.1102       & 0.1308         &     0.1253      & {\ul 0.2230} & 0.1001        & 0.1463            & \textbf{0.2576}                   \\
\hline
\end{tabular}       \\      
\footnotesize{$^*$ The depth of embedding propagation layer of NGCF and LightGCN is the same as LT-HGCF's and equals $4$.}\\
\end{table*}

\subsubsection{Baseline methods}
We compare LT-HGCF with seven methods, where singular value decomposition~(SVD) is a basic linear CF recommendation method, MLP, NeuMF and DeepCF are neural recommendation methods that only utilize user-item interactions, ANR is a neural recommendation method that leverages textual information, NGCF and LightGCN are the state-of-the-art graph-based  methods.

\textbf{SVD}~\cite{zhang2005svd} is often used as a benchmark for recommendation tasks, aims to find the linear model that maximizes the log-likelihood of the rating matrix.

\textbf{MLP}~\cite{he2017ncf} applies matching function learning neural networks in collaborative filtering, which utilizes non-linearity of DNNs for modeling user-item latent structures.

\textbf{NeuMF}~\cite{he2017ncf} combines the linearity of MF (GMF, generalized matrix factorization) and non-linearity of DNNs (MLP) for modelling user–item latent structures.

\textbf{DeepCF}~\cite{deng2019deepcf} combines neural representation learning and neural matching function learning to overcome the limited expressiveness and the weakness in capturing low-rank relations.

\textbf{ANR}\footnote{Note that, ANR is implemented with mean squared error loss to predict ratings, which makes it is not comparable with other ranking methods. We modify the loss function of ANR as BPR loss to make it comparable in this work.}~\cite{chin2018anr} applies an attention mechanism to focus on the relevant parts of comments and estimates aspect-level user and item importance jointly with the aspect-based representation learning.

\textbf{NGCF}~\cite{wang2019ngcf} represents user-item interactions as a bipartite graph and explicitly encodes the collaborative signal in the form of high-order connectivity by performing embedding propagation based on graph convolutional network.

\textbf{LightGCN}~\cite{he2020lightgcn} simplifies the design of NGCF to learn embeddings by linearly propagating, which gains significant improvements over NGCF on the bipartite user-item interaction graph.

\subsubsection{Evaluation Methodology and Metrics}
Following ~\cite{he2017ncf, xue2017dmf, deng2019deepcf}, we adopt the leave-one-out evaluation to evaluate the recommendation performance, i.e., the latest interaction of each user is used for testing and the remaining data for training.
For the methods~\cite{he2017ncf, deng2019deepcf, chin2018anr} with neural matching function learning (including our LT-HGCF), it is too time-consuming to rank all items for every user during evaluation. Thus, we follow the common strategy~\cite{koren2008factorization} that randomly samples $99$ items that are not interacted by the user, and then rank these items with the test one item.
The training and testing data is constructed in advance, and all the evaluated methods are trained and tested by using the same data.
Note that, to maintain a fair comparison, we remove the comments of test data in the pre-processing period.

For the testing data, we utilize Hit Ratio (HR) and Normalized Discounted Cumulative Gain (NDCG) on Top-$10$ and Top-$20$ ranking list as the evaluation metrics, denoted as HR@$10$, NDCG@$10$, HR@$20$ and NDCG@$20$.
The HR intuitively measures whether the test item is present on the Top-$k$ list, and the NDCG accounts for the position of the hit by assigning higher scores to hits at top ranks~\cite{he2017ncf}. A higher value of HR and NDCG means a better performance.

\setlength{\tabcolsep}{9pt}
\begin{table*}[t]
	\centering
	\footnotesize
	\caption{Performance comparison between NGCF, LightGCN and LT-HGCF at different layers.}
	\label{tab:layers}
	\linespread{1.1}\selectfont
\begin{tabular}{c|c|cc|cc|cc}
\hline
\multicolumn{2}{c}{\textbf{Dataset}}                  & \multicolumn{2}{|c}{\textbf{Music}} & \multicolumn{2}{|c|}{\textbf{Beauty}} & \multicolumn{2}{c}{\textbf{Clothing}} \\
\hline
\textbf{Layer \#}                 & \textbf{Method}   & \textbf{HR@20}  & \textbf{NDCG@20} & \textbf{HR@20}  & \textbf{NDCG@20}  & \textbf{HR@20}   & \textbf{NDCG@20}   \\
\hline
\multirow{3}{*}{2 Layer} & NGCF     & 0.7450          & 0.4177           & 0.4453          & 0.2284            & 0.2548           & 0.1026             \\
                                  & LightGCN & 0.7869          & 0.4552           & 0.5070          & 0.2735            & 0.2927           & 0.1287             \\
                                  & LT-HGCF  & 0.8231          & 0.4494           & 0.5858          & 0.2989            & 0.5141           & 0.2396             \\
\hline
\multirow{3}{*}{3 Layer} & NGCF     & 0.7519          & 0.4153           & 0.4473          & 0.2259            & 0.2480           & 0.0982             \\
                                  & LightGCN & 0.7903          & 0.4557           & 0.5147          & 0.2778            & 0.3092           & 0.1384             \\
                                  & LT-HGCF  & 0.8258          & 0.4494           & 0.5997          & 0.3082            & 0.5290           & 0.2467             \\
\hline
\multirow{3}{*}{4 Layer} & NGCF     & 0.7533          & 0.4182           & 0.4489          & 0.2276            & 0.2499           & 0.1001             \\
                                  & LightGCN & 0.7946          & 0.4527           & 0.5195          & 0.2792            & 0.3193           & 0.1463             \\
                                  & LT-HGCF  & 0.8439          & 0.4762           & 0.6068          & 0.3173            & 0.5426           & 0.2576              \\ 
\hline        
\end{tabular}
\end{table*}

\setlength{\tabcolsep}{12pt}
\begin{table*}[h]
	\centering
	\footnotesize
	\caption{Types of textual information and initialization methods. Best performance is in boldface.}
	\label{tab:text-info}
	\linespread{1.1}\selectfont
	\begin{tabular}{c|c|cc|cc|c}
		\hline
		\multirow{2}[0]{*}{\textbf{Dataset}} & \multirow{2}[0]{*}{\textbf{Metric}} & \multicolumn{2}{c|}{\textbf{Textual Information}} & \multicolumn{2}{c|}{\textbf{Initialization Methods}} & \multirow{2}[0]{*}{\textbf{Default}} \\
		\cline{3-6}&       & \textbf{W/O Com.} & \textbf{W/O Des.} & \textbf{W/O Pre.} & \textbf{W/ GloVe} &       \\
		\hline
		\multirow{4}[0]{*}{Music}
		& HR@10 & 0.6504  & 0.6874  & 0.4023  & 0.6318  & \textbf{0.7011} \\
	   	& NDCG@10 & 0.4038  & 0.4327  & 0.2293  & 0.3867  & \textbf{0.4424} \\
   		& HR@20 & 0.8026  & 0.8233  & 0.5589  & 0.7833  & \textbf{0.8439} \\
   		& NDCG@20 & 0.4421  & 0.4672  & 0.2687  & 0.4250  & \textbf{0.4762} \\
		\hline
		\multirow{4}[0]{*}{Beauty}
		& HR@10 & 0.4327  & 0.4398  & 0.2121  & 0.4229  & \textbf{0.4747} \\
		& NDCG@10 & 0.2566  & 0.2610  & 0.1146  & 0.2484  & \textbf{0.2840} \\
		&HR@20 & 0.5691  & 0.5745  & 0.3336  & 0.5593  & \textbf{0.6068} \\
		& NDCG@20 & 0.2895  & 0.2949  & 0.1450  & 0.2823  & \textbf{0.3173} \\
		\hline
		\multirow{4}[0]{*}{Clothing}
		& HR@10 & 0.3432  & 0.3422  & 0.1106  & 0.3203  & \textbf{0.3896} \\
		& NDCG@10 & 0.1872  & 0.1865  & 0.0521  & 0.1700  & \textbf{0.2191} \\
		& HR@20 & 0.5025  & 0.4995  & 0.2093  & 0.4826  & \textbf{0.5426} \\
		& NDCG@20 & 0.2262  & 0.2252  & 0.0765  & 0.2109  & \textbf{0.2576} \\
		\hline
	\end{tabular}
\end{table*}

\subsection{Performance Comparison~(RQ1)}
Table~\ref{tab:comparison} shows the summarized results of our experiments on the three datasets in terms of metrics including HR@$10$, NDCG@$10$, HR@$20$, and NDCG@$20$. From the results, we have the following key observations:

\begin{itemize}
\item Compared with the linear method SVD, the non-linear neural network methods perform better. DeepCF and NeuMF\footnote{Note that, we do not utilize the pre-train components to initialize DeepCF, which may result in the performance gap between DeepCF and NeuMF.}, which combine representation learning and matching function learning, always perform better than MLP. Moreover, among these methods that only use user-item interaction information, graph-based LightGCN performs the best.

\item For the \textbf{Music} dataset that is less sparse, the textual ANR cannot compete with most of the methods. However, with the increase of data sparsity, the strength of ANR is shown out. On \textbf{Clothing}, ANR outperforms other  methods significantly, which demonstrates the importance of utilizing textual information to alleviate the negative effects of data sparsity.

\item By exploiting pre-trained SBERT embeddings to initialize the text nodes on the heterograph and combining a light RGCN-based representation learning with matching function learning to learn the preference of users, our LT-HGCF consistently outperforms all the evaluated baselines and obtains remarkable improvements.
\end{itemize}

Since the depth of embedding propagation layer is important for the graph-based methods, we compare NGCF, LightGCN, and LT-HGCF in different depths at HR@$20$ and NDCG@$20$.
As shown in Table~\ref{tab:layers}, in general, with the growth of layer number from $2$ to $4$, the performance of these methods all improves,  and the best performance is obtained with $4$ layers. Less or more hidden layers would decrease the model performance.  It is straightforward that less layers are hard to fit the model completely, leading to the classic underfitting problem, while more layers may overfit the model. Moreover, no matter how the layer number varies, the performance of LT-HGCF is always the best, which further verifies the superior performance of our method.

\setlength{\tabcolsep}{0.5pt}
\begin{table*}[h]
 \centering
 \footnotesize
 \caption{The ablation experiment for the main components. Best performance is in boldface.}
 \label{tab:ablation}
 \linespread{1.1}\selectfont
\begin{tabular}{c|c|c|c|cc|c|c|c}
\hline
\textbf{Dataset}          & \textbf{Metric} & \textbf{W/ Weight} & \textbf{W/ Active.} & \textbf{W/ Self-con.} & \textbf{W/O Layer Com.} & \textbf{W/O Init. Resid.} & \textbf{W/ GCN} & \textbf{Default} \\
\hline
\multirow{4}{*}{Music}    & HR@10           & 0.6576             & 0.6807              & 0.7002                & 0.6775                  & 0.6890                    & 0.6683          & \textbf{0.7011}  \\
                          & NDCG@10         & 0.4051             & 0.4214              & 0.4372                & 0.4253                  & 0.4237                    & 0.4073          & \textbf{0.4424}  \\
                          & HR@20           & 0.8094             & 0.8246              & 0.8396                & 0.8273                  & 0.8381                    & 0.8190          & \textbf{0.8439}  \\
                          & NDCG@20         & 0.4436             & 0.4579              & 0.4714                & 0.4609                  & 0.4616                    & 0.4449          & \textbf{0.4762}  \\
\hline
\multirow{4}{*}{Beauty}   & HR@10           & 0.4244             & 0.4486              & 0.4630                & 0.4469                  & 0.4655                    &0.4442          & \textbf{0.4747}  \\
                          & NDCG@10         & 0.2466             & 0.2661              & 0.2795                & 0.2644                  & 0.2789                    & 0.2594          & \textbf{0.2840}  \\
                          & HR@20           & 0.5556             & 0.5878              & 0.5946                & 0.5847                  & 0.6029                    &0.5781          & \textbf{0.6068}  \\
                          & NDCG@20         & 0.2798             & 0.2998              & 0.3116                & 0.2992                  & 0.3135                    & 0.2928          & \textbf{0.3173}  \\
\hline
\multirow{4}{*}{Clothing} & HR@10           & 0.3245             & 0.3549              & 0.3835                & 0.3813                  & 0.3776                    &0.3342          & \textbf{0.3896}  \\
                          & NDCG@10         & 0.1735             & 0.1954              & 0.2183                & 0.2088                  & 0.2131                    & 0.1822          & \textbf{0.2191}  \\
                          & HR@20           & 0.4831             & 0.5141              & 0.5315                & 0.5422                  & 0.5320                    & 0.4824          & \textbf{0.5426}  \\
                          & NDCG@20         & 0.2123             & 0.2354              & 0.2556                & 0.2495                  & 0.2520                    & 0.2194          & \textbf{0.2576} \\
\hline
\end{tabular}
\label{tab:addlabel}%
\end{table*}%

\subsection{Effect of Textual Information~(RQ2)}
To further figure out the impact of the textual information, in this part, we conduct experiments and discuss the results from different types of textual information and  different methods for the initialization of text nodes in the heterograph.

As shown in Figure~\ref{fig:graph}, the heterograph we used contains two types of textual information, descriptions and comments. Therefore, we consider three possible cases, without comments (W/O Com. for short), without descriptions (W/O Des. for short), and with all textual information, i.e., Default. 

It can be observed from Table~\ref{tab:text-info} that without comments or descriptions result in poor performance, demonstrating that the textual information contains rich knowledge of users and items. Moreover,  compared with the case that only exploits one kind of textual information, combining both them can always achieve better performance.
For the \textbf{Music} dataset, comments contribute more than descriptions on performance improvement. However, with the increase of data sparsity, descriptions play a nearly equal important role.
This may be due to that the comments have a more positive correlation with the interactions, as compared to descriptions. Moreover, the growth of interaction sparsity also brings a negative effect on the introduction of comments.

In this work, we propose to initialize the text nodes' embeddings in the heterograph with the non-contextual method GloVe (W/ GloVe for short) and the contextual sentence encoding method SBERT, respectively.
The right part of Table~\ref{tab:text-info} shows that compared to the method without pre-training (i.e., randomly initialize the text nodes' embeddings, W/O Pre. for short),  the methods that are initialized with advanced NLP techniques can enrich the knowledge of embeddings and thus gain a better performance. Further, compared with the non-contextual method GloVe, the contextual method SBERT (the Default) always performs better.

\subsection{Ablation Experiment~(RQ3)}

Our LT-HGCF model discards the weight of the embedding network, and the third column in Table~\ref{tab:ablation} demonstrates that the addition of weight matrices (W/ Weight for short) will lead to a performance degradation. The activation functions of embedding and predictive network are removed in LT-HGCF, and the fourth column in Table~\ref{tab:ablation} shows that, after adding the LeakyReLU activation function on both networks (W/ Active. for short), the model performance decreases.
According to the analysis of LightGCN, the weight and activation function may increase the difficulty for training, which results in a performance degradation.
In the middle part of Table~\ref{tab:ablation}, we compare the default setting with the one that adds self-connection and removes the layer combination operation (W/ Self-co. for short), and the one that only abandons the layer combination operation (W/O LaCo. for short). From the fifth and sixth columns, we can conclude that the layer combination operation can improve the performance of our model and is more effective than self-connection.
As shown in the seventh column in Table~\ref{tab:ablation}, without the variation of the initial residual (W/O Resid. for short), the performance of LT-HGCF will degenerate, demonstrating its effectiveness. Moreover, the  hyper-parameter experiment latter on the depth of propagation layer can show its strength in alleviating the over-smoothing problem.
To verify the necessity of utilizing heterogeneous representation learning methods on heterographs, we change the embedding method to the modified GCN (W/ GCN for short).
Note that, for fair competition, the modified GCN method also applies the same modifications as our method.
It can be observed from the the results in Table~\ref{tab:ablation} that it is harmful to degenerate the heterograph learning process as a homogeneous one.

\begin{figure*}[t]
	\centering
	\subfloat[\scriptsize{}]{
		\begin{minipage}[t]{0.47\textwidth}
			\centering
			\includegraphics[width=\linewidth]{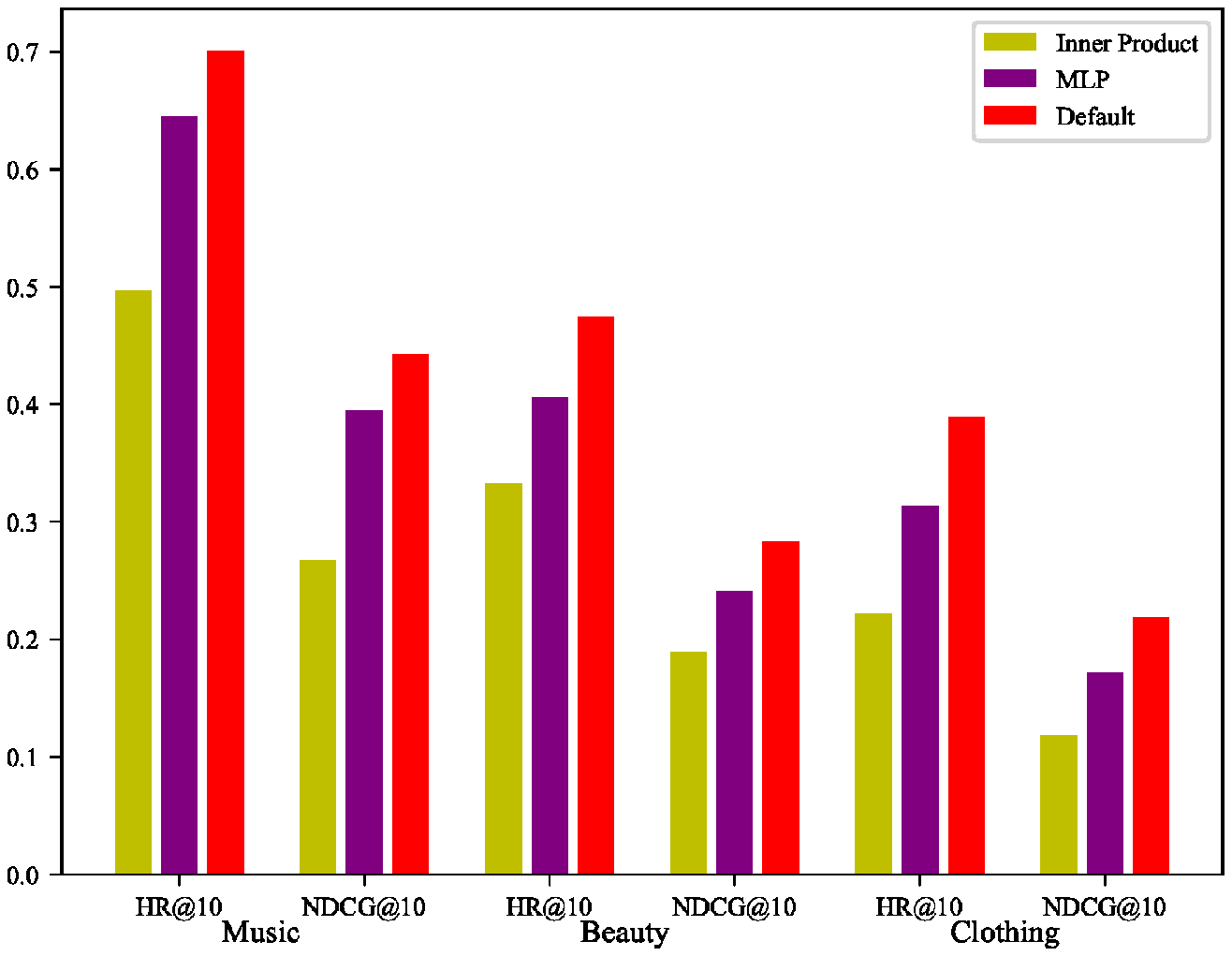}
		\end{minipage}
	}
	\subfloat[\scriptsize{}]{
		\begin{minipage}[t]{0.47\textwidth}
			\centering
			\includegraphics[width=\linewidth]{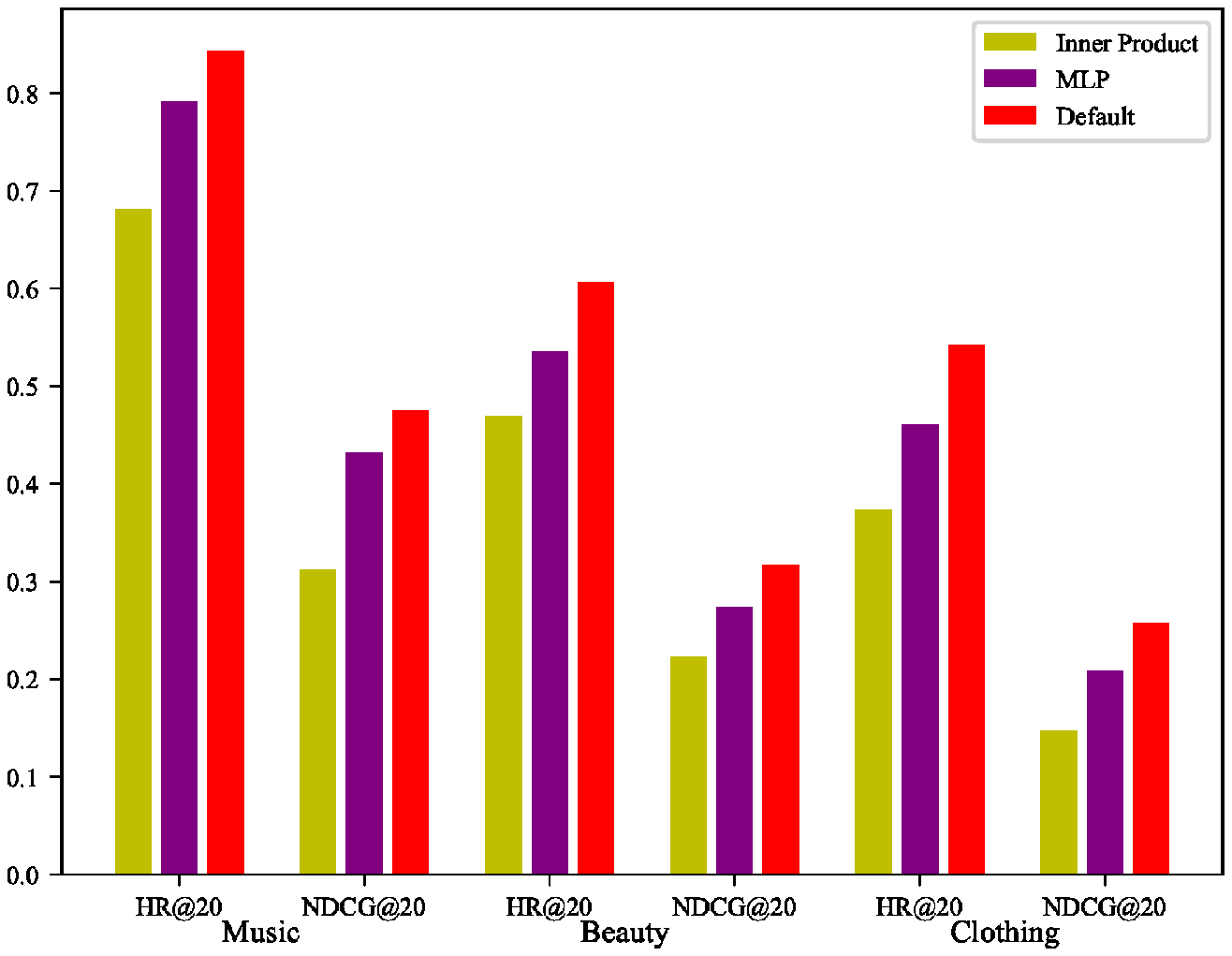}
		\end{minipage}
	}
	\caption{The different types of matching functions \emph{w.r.t.}  (a) the performance on Top-$10$; (b) the performance on Top-$20$.}
	\label{fig:p0}
\end{figure*}

From Figure~\ref{fig:p0}, we can observe that using neural matching function learning can achieve better performance, and our predictive network that combines graph-based representation learning with matching function learning can always gain the best performance.
Compared with the simple inner product, the neural network can deal with the complex semantics obtained from the textual information better, and thus result in a significant performance improvement.

\begin{figure*}[t]
	\centering
	\subfloat[\scriptsize{}]{
		\begin{minipage}[t]{0.32\textwidth}
			\centering
			\includegraphics[width=\linewidth]{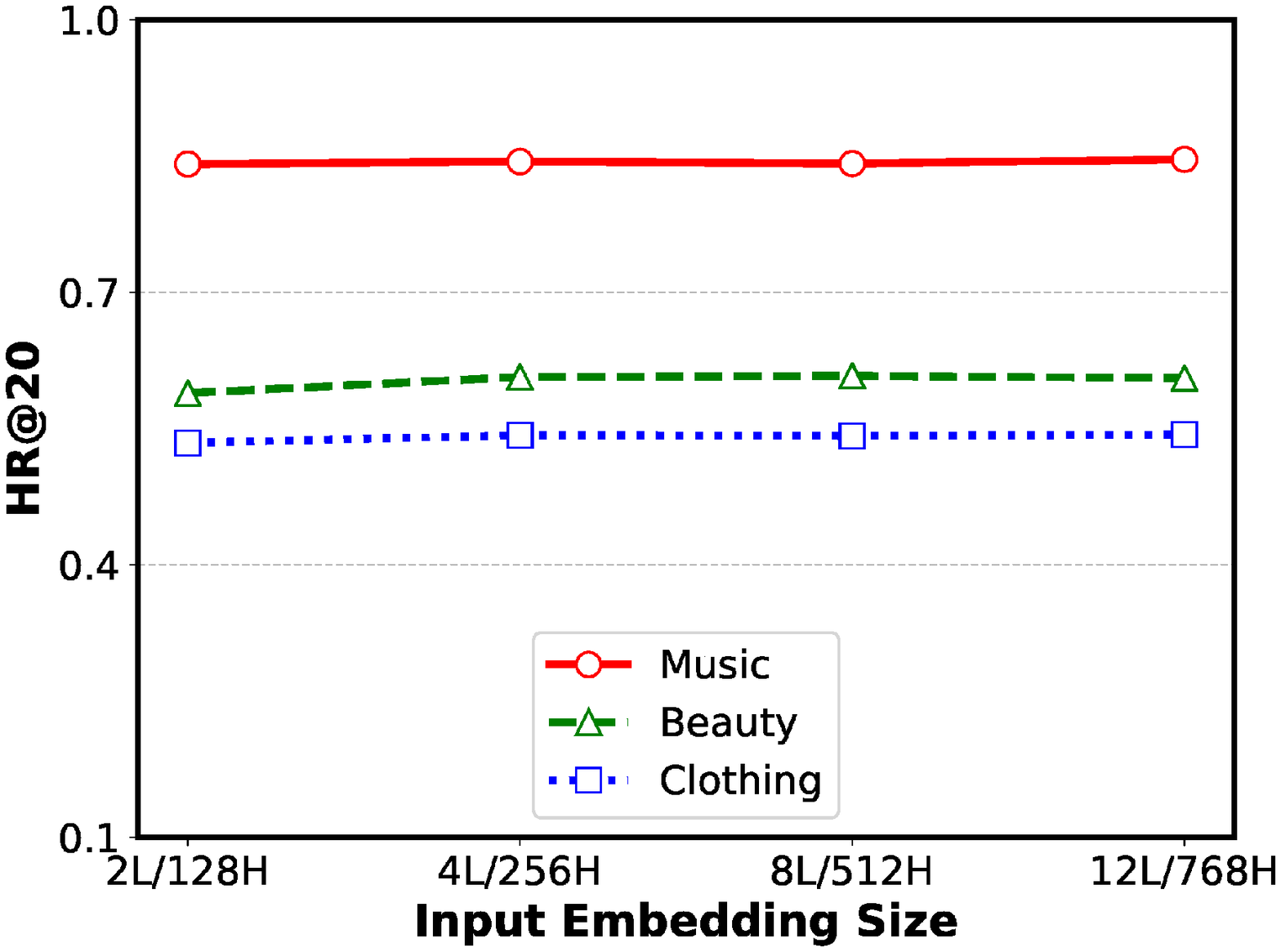}
		\end{minipage}
	}
	\subfloat[\scriptsize{}]{
		\begin{minipage}[t]{0.32\textwidth}
			\centering
			\includegraphics[width=\linewidth]{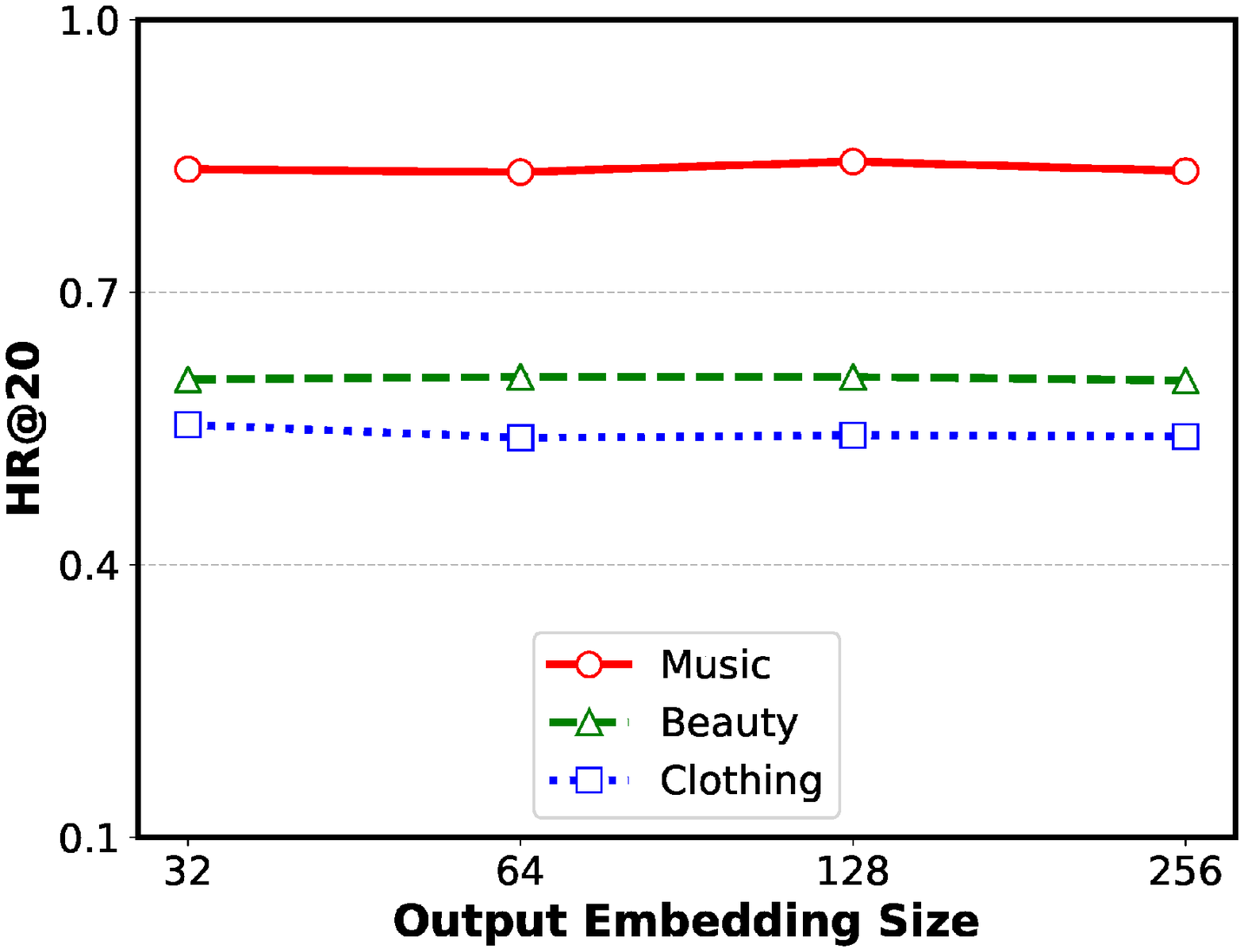}
		\end{minipage}
	}
	\subfloat[\scriptsize{}]{
		\begin{minipage}[t]{0.32\textwidth}
			\centering
			\includegraphics[width=\linewidth]{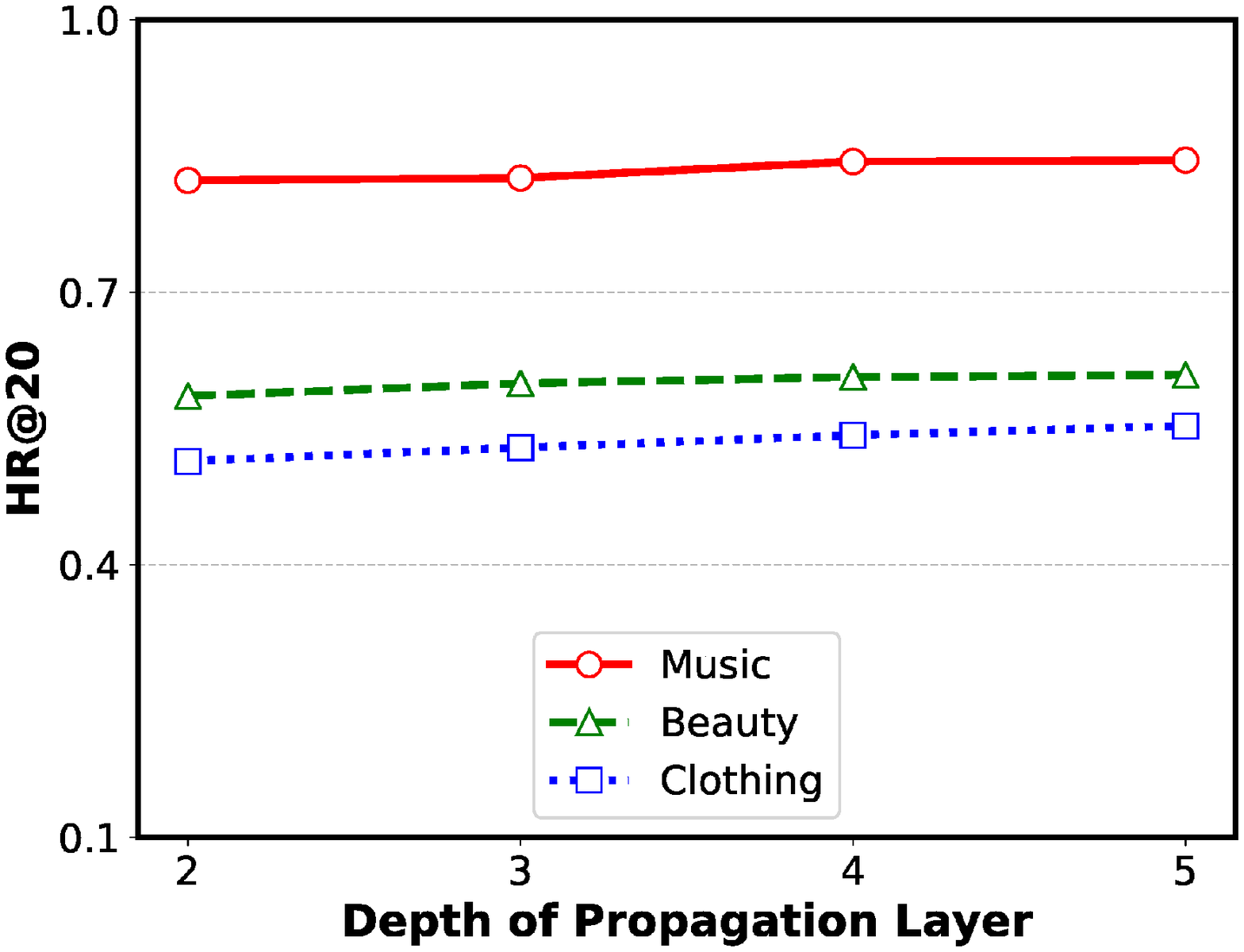}
		\end{minipage}
	}\\
	\caption{Performance of HR@20 \emph{w.r.t.}  (a) the input size of embedding network; (b) the output size of embedding network; (c) the depth of embedding propagation layer.}
	\label{fig: p1}
\end{figure*}

\begin{figure*}[t]
	\centering
	\subfloat[\scriptsize{}]{
		\begin{minipage}[t]{0.32\textwidth}
			\centering
			\includegraphics[width=\linewidth]{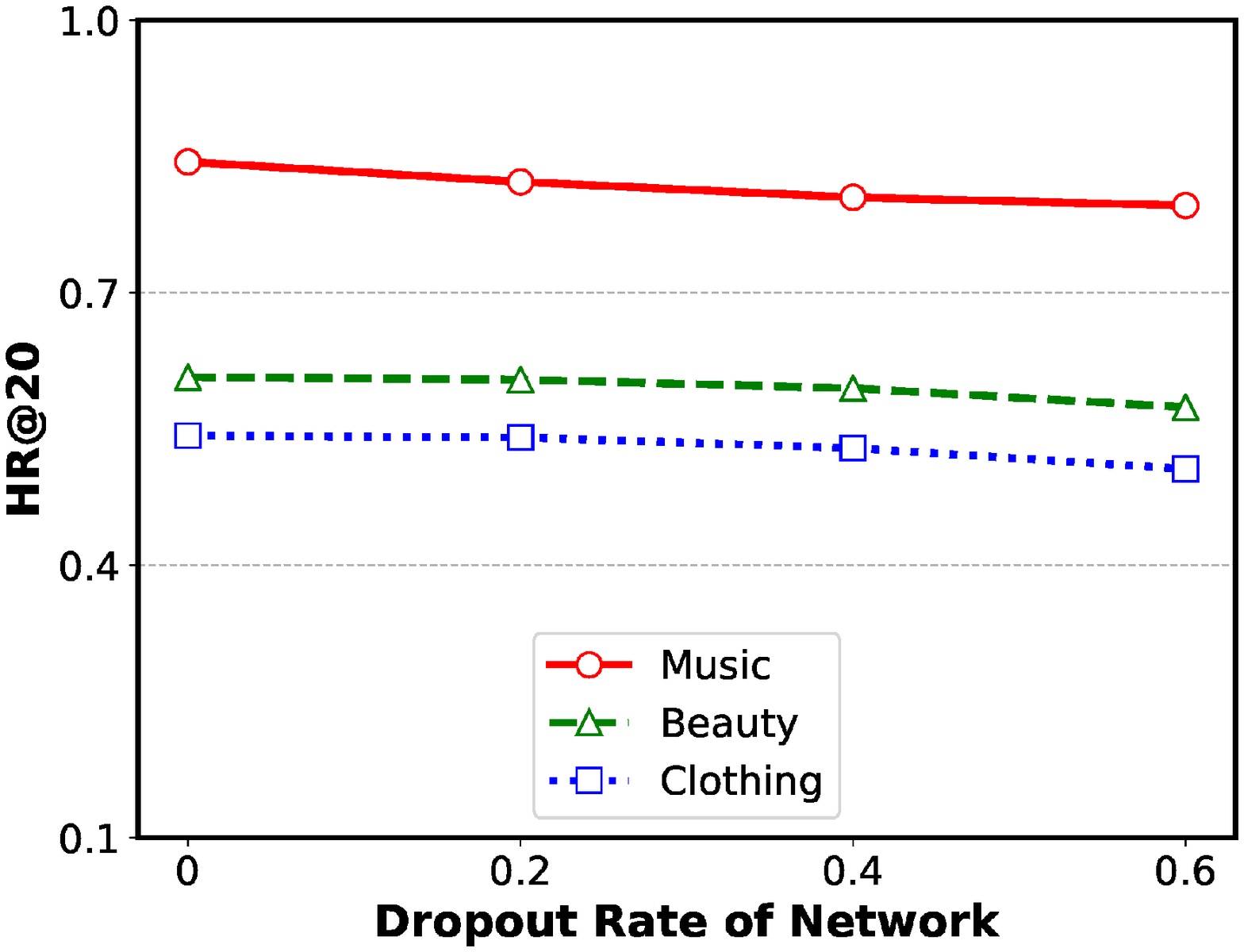}
		\end{minipage}
	}
	\subfloat[\scriptsize{}]{
		\begin{minipage}[t]{0.32\textwidth}
			\centering
			\includegraphics[width=\linewidth]{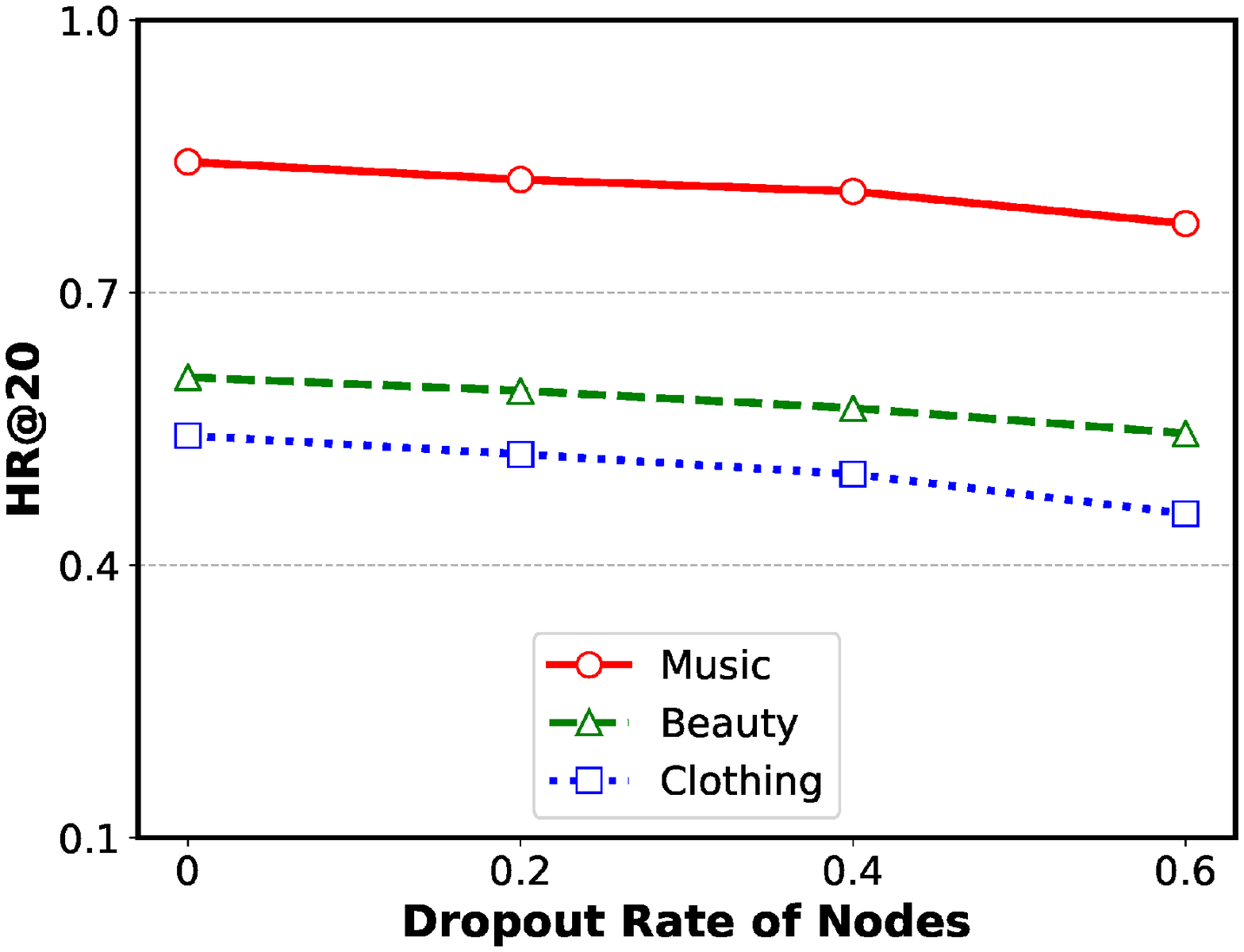}
		\end{minipage}
	}
	\subfloat[\scriptsize{}]{
		\begin{minipage}[t]{0.32\textwidth}
			\centering
			\includegraphics[width=\linewidth]{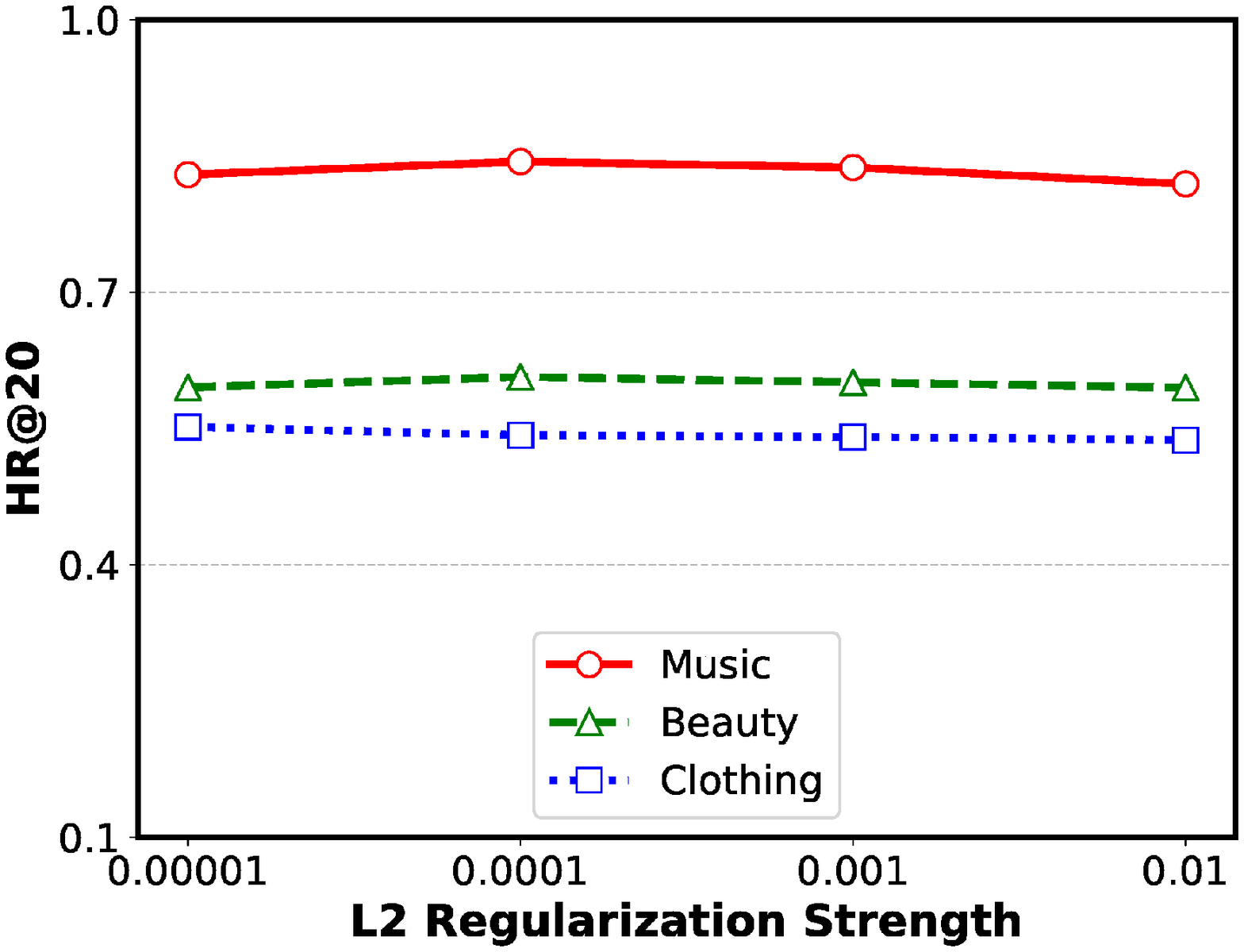}
		\end{minipage}
	}\\
	\caption{Performance of HR@20 \emph{w.r.t.}  (a) the dropout rate of network; (b) the dropout rate of nodes; (c) the $L_2$ regularization strength.}
	\label{fig: p2}
\end{figure*}

\begin{figure*}[t]
	\centering
	\subfloat[\scriptsize{}]{
		\begin{minipage}[t]{0.32\textwidth}
			\centering
			\includegraphics[width=\linewidth]{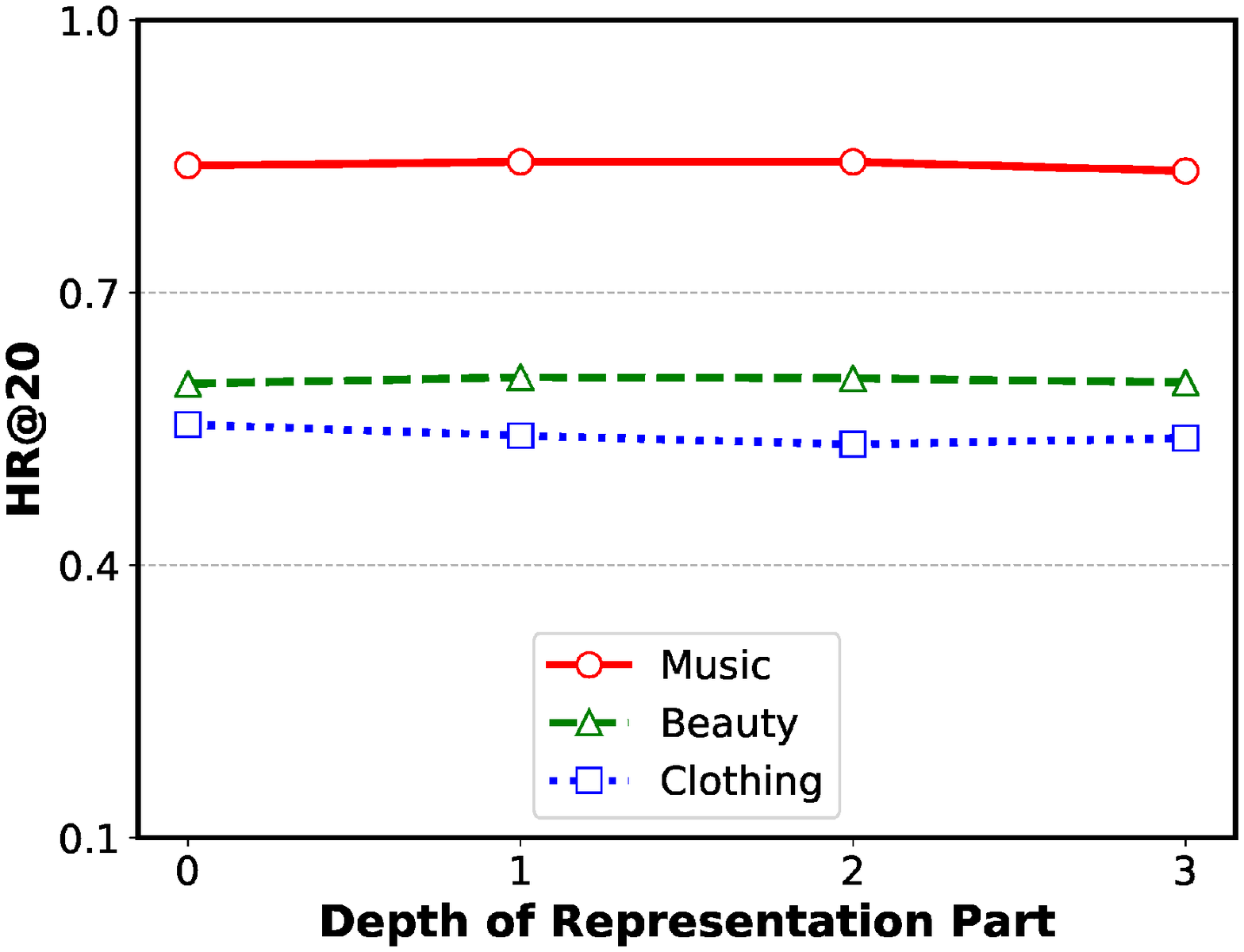}
		\end{minipage}
	}
	\;\;\;\;\;\;
	\subfloat[\scriptsize{}]{
		\begin{minipage}[t]{0.32\textwidth}
			\centering
			\includegraphics[width=\linewidth]{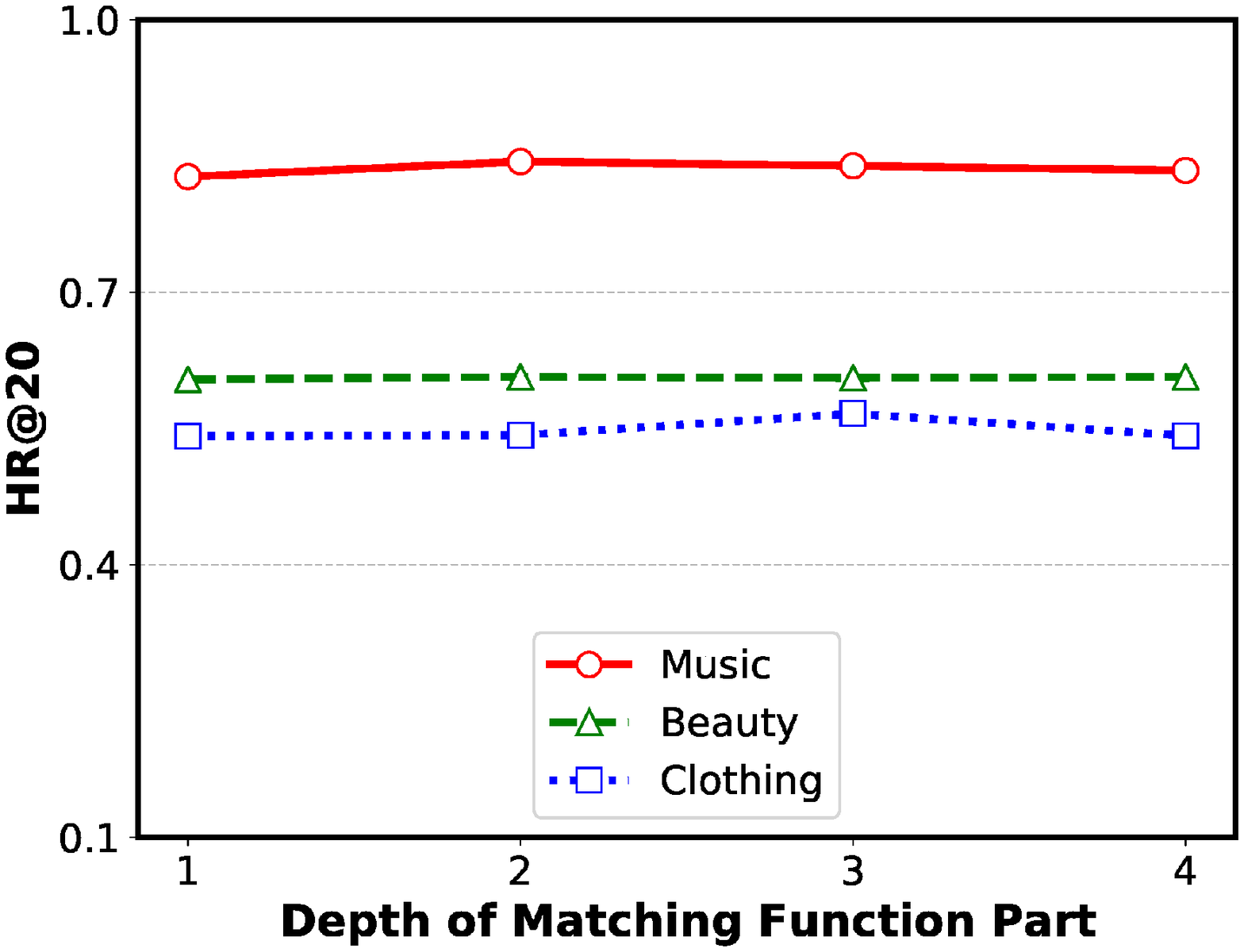}
		\end{minipage}
	}\\
	\caption{Performance of HR@20 \emph{w.r.t.}  (a) the depth of representation network; (b) the depth of matching function network..}
	\label{fig: p3}
\end{figure*}

\subsection{Hyper-parameter Sensitivity~(RQ4)}
We select several important parameters~(input and output size of embedding network, dropout rate of the network and the nodes, depth of embedding propagation layer, depth of predictive network, and the $L_2$ regularization strength $\lambda$) to analyze their effects on the performance of our method. Since HR@10 exhibits a similar performance trend as HR@20, for simplicity, we only present HR@20 of each dataset in the followings.

\subsubsection{Input and output size of embedding network.}
Lulia \emph{et al.}~\cite{turc2019small_bert} released a couple of pre-trained miniature BERT models, which have smaller transformer layers~(L) and hidden embedding sizes~(H). The horizontal axis of Figure~\ref{fig: p1}~(a) gives four combinations of L and H, from left to right are BERT-Tiny, BERT-Mini, BERT-Medium, and BERT-Base. It is clear to see that, the larger input size means more knowledge, which can result in better performance.

Figure~\ref{fig: p1}~(b) shows that the performance of our method does not improve with the increase of the output size of the embedding network. The reason might be that the useful knowledge can be contained within $32$ dimensions.

\subsubsection{Depth of embedding propagation layer.}
Figure~\ref{fig: p1}~(c) shows that with the increase of the embedding propagation layer number, the performance of our model also increases, which is much different from the trends of NGCF and LightGCN.
The reason might be that the introduction of the variation of the initial residual directly adds the first-order embedding to deeper layers, which alleviates the over-smoothing problem to some extent and improves the model performance.

\subsubsection{Dropout rate of network and nodes.}
In Figure~\ref{fig: p2}~(a) and (b), when the dropout rate of network and nodes ranges from $0$ to $0.6$, the performance of our model decreases accordingly, which has a similar trend with that of LightGCN.
That is because the model has discarded the weight matrices on the embedding part and the activation function, which makes it simple enough, the dropout rate is therefore superfluous.

\subsubsection{The $L_2$ regularization strength.}
Figure~\ref{fig: p2}~(c) shows that with the increase of the $L_2$ regularization strength $\lambda$, the performance of our model first rises and then falls. It achieves the best performance when $\lambda=1 \times 10^{-4}$. It can be  observed that too small $\lambda$ may trap the model in a local optimum, while too large $\lambda$ may miss the best parameters.

\subsubsection{Depth of predictive network.}
As shown in Figure~\ref{fig:joint-net}, in our predictive network, the depth of the fusion layer is set to $1$.
Figure~\ref{fig: p3} (a) and (b) show the performance of our method with varying depths of MLP layers for representation learning and  matching function learning, respectively.

For the representation learning part, the best performance is achieved when the depth is $0$ or $1$. Representation learning is applied to get the proper representations of users and items, this work has been completed by the graph convolutional network, so only one layer is enough to adjust the representations. For the matching function learning part, proper depth of the network can improve the model learning ability, but too more layers will make the model difficult to train.

\section{Conclusion}
\label{sec: conclusion}
In this paper, to exploit the rich preference knowledge in textual information, we first build a heterograph with comment and description nodes, and then utilize the pre-trained NLP models (GloVe or SBERT) to initialize the embeddings of these text nodes.
On the constructed heterograph, based on the modified RGCN, our LT-HGCF model propagates information among nodes.
As for the modified RGCN, we discard feature transformation and non-linear activation of it to simplify the model while achieving better performance.
Besides, we adopt the layer combination in LightGCN and the initial residual in GCNII to further strengthen the representation ability of RGCN.
By the message propagation among neighbors, the knowledge in text nodes can be aggregated to the representations of users and items. Moreover, we find that the matching function used by most graph-based representation learning methods is the inner product, which is insufficient to capture the complex semantics.
To better fit the semantics, inspired by GraphRec and DeepCF, we propose a framework that can combine our graph-based representation learning method with the neural matching function learning, and train the RGCN-based embedding network and predictive network jointly. We conduct extensive experiments on three public datasets, the results verify the superior performance of  LT-HGCF  over several baselines.

For future work, we intend to study whether it is possible to incorporate our proposed method with reinforcement learning and transfer learning to meet the different needs of recommender systems.


\section*{Acknowledgment} We would like to thank the editors and anonymous reviewers for their valuable suggestions and comments, which helped improve this paper considerably. The work was partially supported by the National Natural Science Foundation of China under Grant No.61672252.

\bibliography{mybibfile}

\end{document}